\documentclass[12pt]{article}
\usepackage{mathrsfs}
\usepackage{amsmath,amssymb}
\usepackage{epsfig}
\usepackage{relsize}
\usepackage{bbm}
\usepackage{graphicx}
\usepackage{float}
\usepackage{color}
\usepackage{enumitem}

\def\({\left(}
\def\){\right)}

\newcommand{\Y}{\mathcal Y}
\newcommand{\E}{\mathcal E}

\def\tH{\widetilde{\mathbf H}}

\def\sl(2){\alg{sl}(2)}

\def \cs {{\alg s}}

\def\be{\begin{equation}}
\def\ee{\end{equation}}

\newcommand{\bea}{\begin{eqnarray}}
\newcommand{\eea}{\end{eqnarray}}

\newcommand{\bei}{\begin{itemize}}
\newcommand{\eei}{\end{itemize}}

\newcommand{\bee}{\begin{enumerate}}
\newcommand{\eee}{\end{enumerate}}

\def\br {\bar{\rho}}
\def\a {\alpha}

\def\s {\sigma}

\def\g {\gamma}
\def\om {\omega}

\def\la{\label}
\def\e{\epsilon}
\def\ov{\over}
\def\tr{{\rm tr}}

\newcommand{\alg}[1]{\mathfrak{#1}}
\newcommand{\su}{\alg{su}}

\newcommand{\AdS}{{\rm  AdS}_5\times {\rm S}^5}

\newcommand{\bem}{\left (\begin{matrix}}
\newcommand{\eem}{\end{matrix} \right )}

\def\ka{\kappa}
\def\lam{\lambda}

\def\V{{\mathscr V}}
\def\I{{\cal I}}

\def\cex{{_{\cal C}}}
\def\sucex{\su(2|2)\cex}

\def\cs{\mathbf s}

\def\cm{\mathbf c}
\def\cd{{\mathbf c}^\dagger}
\def\n{\mathbf n}

\def\n{\mathbf n}

\def\bstar{\,\bar{\star}\,}
\def\hstar{\,\hat{\star}\,}
\def\cstar{\,\check{\star}\,}

\def\uh{ \mathfrak u}
\def\gh{ \mathfrak g}

\def\bs{\beta_s}
\def\bc{\beta_c}
\def\bd{\beta_\Delta}

\def\cK{{\cal K}}
\def\mK{{\mathscr K}}

\def\tY{{\widetilde Y}}
\def\tB{{\widetilde B}}
\def\tmu{{\tilde \mu}}
\def\tf{{\tilde f}}

\def\tj{{\it{t-J }}}

\newcommand{\bean}{\begin{eqnarray*}}
\newcommand{\eean}{\end{eqnarray*}}

\usepackage{hyperref}
\hypersetup{
plainpages=false
}

\begin{document}



\null\vskip-40pt
 \vskip-5pt \hfill
\vskip-5pt \hfill {\tt\footnotesize
TCD-MATH-11-14}
 \vskip-5pt \hfill {\tt\footnotesize
HMI-11-05}

\vskip 1cm \vskip0.2truecm
\begin{center}
\begin{center}
\vskip 0.8truecm {\Large\bf Hubbard-Shastry lattice models
}
\end{center}

\renewcommand{\thefootnote}{\fnsymbol{footnote}}

\vskip 0.9truecm
Sergey Frolov\footnote[1]{ Correspondent fellow at
Steklov Mathematical Institute, Moscow.} \  and Eoin Quinn\footnote[2]{emails: 
frolovs@maths.tcd.ie, epquinn@gmail.com}
 \\
\vskip 0.5cm

{\it School of Mathematics and Hamilton Mathematics Institute, \\
Trinity College, Dublin 2,
Ireland}

\end{center}
\vskip 1cm \noindent\centerline{\bf Abstract} \vskip 0.2cm 

We consider two lattice models  for strongly correlated electrons which are exactly-solvable in one dimension. 
Along with the Hubbard model and the $\su(2|2)$ spin chain, these are the only parity-invariant  models that can be obtained from Shastry's R-matrix.
One exhibits itinerant ferromagnetic  behaviour, 
 while for the other the electrons form bound pairs and at half-filling the model becomes insulating.
We derive the TBA equations for the models,  analyze them at various limits, and in particular obtain zero temperature phase diagrams.
Furthermore we consider extensions of the models, which reduce to  the 
Essler-Korepin-Schoutens model in certain limits.

\newpage

\tableofcontents

\renewcommand{\thefootnote}{\arabic{footnote}}
\setcounter{footnote}{0}

\numberwithin{equation}{section}



\section{Introduction and Summary  }

There exist a range of materials whose behaviour is strongly correlated and for which a proper understanding has yet to be established. Prominent in this list are the high-T$_{c}$ superconductors, itinerant ferromagnets and heavy fermion systems. In this paper we present models which  we believe capture some of the features of the first two, and will also briefly comment on a possible connection to the third. 

The complete description of a solid is a complex many body problem. Simplifications and exact methods are crucial for progress to be made. In particular, exactly solvable models provide a bedrock on which more realistic models can be constructed and examined.  The most important such model is that of free particles. It can be solved in any dimension for both bosons and fermions. The physics of the majority of metals is understood by considering the effects of interaction perturbatively on an exact free solution.

Strongly correlated materials however are those which cannot be understood by reducing the complexity to a non-interacting picture. The collective behaviour of the constituents can drive the material into unexpected phases and interacting exactly solvable models are invaluable for getting insight into such effects. The Hubbard model \cite{Hubbard} is a prominent example. Indeed it has become a cornerstone of investigations into strongly correlated electronic behaviour. The model is  integrable in one dimension, and describes a   Mott insulator \cite{Mott}, which can be seen through its exact solution \cite{LiebWu}. The book \cite{book} provides an extensive review of the one-dimensional Hubbard model. 

\medskip

The primary focus of this paper is to investigate two models closely related to the Hubbard model, we call them the A-model and the B-model. Like the Hubbard model they describe spin-1/2 electrons interacting on a lattice and in one dimension can be obtained from Shastry's R-matrix \cite{Shastry}. For this reason we refer to them as Hubbard-Shastry models, and by construction they are  exactly solvable in one dimension. 
Normally an R-matrix depends on two spectral parameters through their difference, and is thus invariant under shifts of the spectral parameters. Shastry's R-matrix is special however as it is of a non-difference form and so the actual value of the spectral parameters are important. This was noted in particular by \cite{swad,uswad}, who used this extra variable to construct a one-parameter extension of the Hubbard model. Their extension is hermitian for purely imaginary values of this parameter and is in fact also hermitian along another line for complex values of the parameter. The model is parity invariant however only for special distinct values of the parameter and these correspond, in addition to the Hubbard model, to the A- and B-models, and the $\su(2|2)$ spin chain.

The A- and B-models have symmetries that are similar to those of the Hubbard model but the phases they exhibit are different. 
The A-model ground state contains spin-polarised electrons and thus  describes  itinerant ferromagnetism.  For the B-model the electrons form bound pairs and at half-filling it becomes insulating.  This is reminiscent  of the behaviour of the supersymmetric $t$-$J$ model \cite{stJ}, and has relevance for the modelling of high temperature superconductivity.

The A- and B-models are more involved than the Hubbard model. Both have correlated hopping terms, where the magnitude of the hopping parameters depend on the occupation of the sites by other electrons, and take into account processes such as nearest neighbour Coulomb interaction, spin exchange of two neighbouring electrons of opposite spin, and the pair hopping of two electrons from one site to a neighbouring site. The relative contributions of these terms are strongly constrained by the integrability of the models. Like the Hubbard model they both have one free coupling constant $\uh$ after one normalises the Hamiltonian so that hopping amplitudes are of order 1.

The A-model does not have a free fermion limit, rather the weak coupling $\uh\rightarrow 0$ limit is a model of graded permutations, the $\su(2|2)$-spin chain \cite{su22}. In the strong coupling limit the Hubbard on-site interaction is the dominant term in the Hamiltonian. The model behaves as a ferromagnetic $t$-$J$ model however, in contrast with the antiferromagnetic behaviour of the  Hubbard model in this limit.

The B-model does describe free spin-1/2 fermions in the weak coupling limit. As the coupling is increased the hopping becomes correlated and all of the interactions mentioned above begin to contribute. The Hubbard interaction appears in the Hamiltonian but there is no range of $\uh$ for which it is dominant.  In the strong coupling limit all terms are of the same order of magnitude, and the model appears to be equivalent to the Essler-Korepin-Schoutens (EKS) model \cite{EKS} with coupling $U=4$. 
Furthermore, the model possesses  a hidden $\su(2|2)$ fermionic symmetry for arbitrary coupling. 

\medskip

The integrability of the models is used to study their exact solution in one dimension. As for the Hubbard model, this can be done for arbitrary chemical potential $\mu$ and applied magnetic field $B$. Our formalism allows us to study the Hubbard, A- and B-models in parallel. 
We restrict our attention to the parameter region $\mu\leq 0$ and $B\geq 0$ (results for the other regions can be obtained using symmetries), here the electron number density ranges from 0 (empty lattice) to 1 (half-filling) and the magnetisation is non-negative. 
The TBA equations of the models are derived and  we analyze them for various limits.
Particular focus is paid to obtaining the zero temperature phase diagram in the thermodynamic limit, as this provides the most insight into their properties. Diagonalising the models using the Bethe ansatz technique, taking the  thermodynamic limit, and going to zero temperature, the ground states are found to be composed of\footnote{ In the conventions of \cite{book} a $w$-particle is a $\Lambda$-string of length 1, 
a $1|vw$-string is a $k$-$\Lambda$ string of length 1, and
$y$-particles could have been called $k$-particles.}
\begin{itemize}
\item $y$-particles: spin-up electrons,
\item $w$-particles: each change the spin of the state by -1,
\item $1|vw$-strings: each is a bound state of one spin-up and one spin-down electron.
\end{itemize}
It is well established \cite{book} that for the Hubbard model there are no $1|vw$-strings in the ground state. In contrast for the B-model there are no $w$-particles, and with no applied magnetic field there are only $1|vw$-strings. The ground state of the  A-model is composed only of $y$-particles. The phase diagrams for the A- and B-models are presented in Figure \ref{phasediags}. The phases identified are: I. empty band, II. partially filled and spin polarised band, III. half-filled and spin-polarised band, IV. partially filled and partially spin polarised band, V. half-filled and partially spin polarised band.

The ground state of  the A-model is spin polarised as it contains only spin-up electrons for $B>0$. The electron number density increases for increasing chemical potential, the lines separating phases I and II and phases II and III depend on the coupling constant respectively as $\mu_1=-2-2\sqrt{1+\uh^2}-B$ and $\mu_2 =2-2\sqrt{1+\uh^2}-B$. 
For $B<0$ the phase diagram is mirrored except that here the ground state is polarised with spin-down electrons. Thus we see that the A-model, which describes itinerant electrons on a one-dimensional lattice, is ferromagnetic.

\begin{figure}[t]
\includegraphics[width=\linewidth]{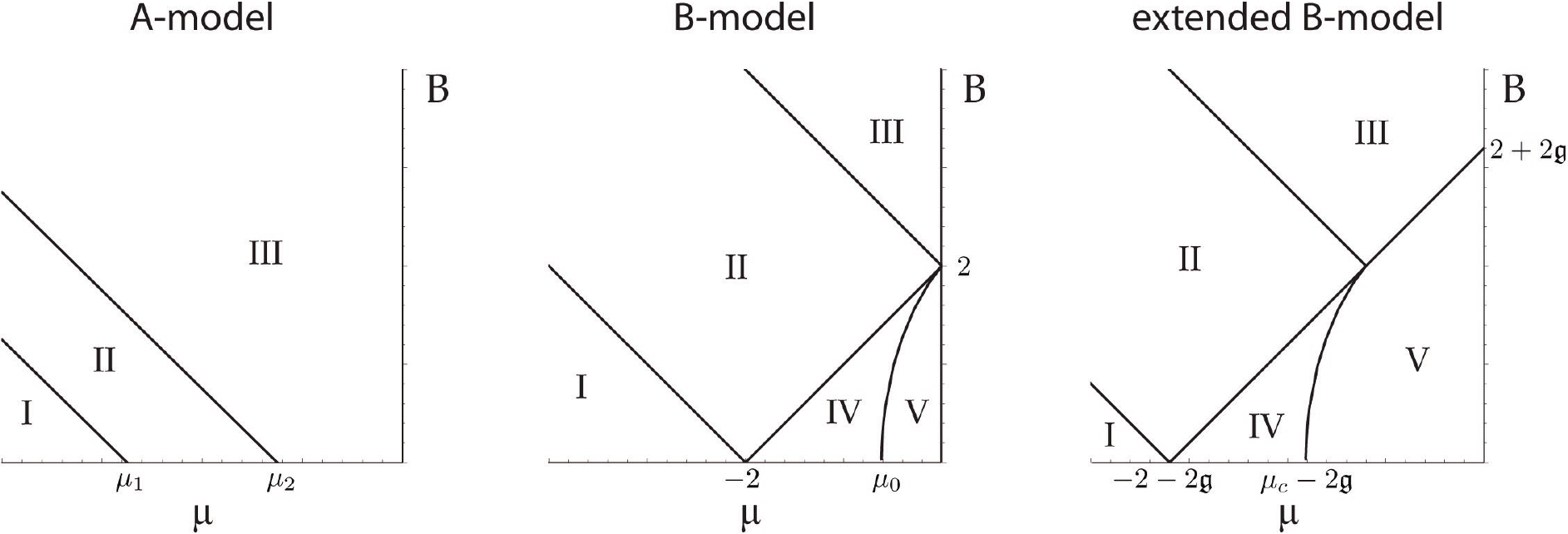}
\vspace{-0.5cm}
\caption{{\small Zero temperature phase diagrams in the $\mu B$-plane  for the A- and B- and extended B-models. The phases identified are: I. empty band, II. partially filled and spin polarised band, III. half-filled and spin-polarised band, IV. partially filled and partially spin polarised band, V. half-filled and partially spin polarised band. The dependence of $\mu_0,\mu_1,\mu_2$ on $\uh$ is given in the text, $\mu_c=-2+2\log 2$, and $\gh$ is the coupling constant of the extended B-model. }}
  \la{phasediags}
\end{figure}

The phase diagram of the B-model is richer. In the weak coupling limit the point  $\mu_0$, whose dependence on $\uh$ is given by eq.\eqref{mu0}, approaches 0, phase V shrinks to a portion of the $B$-axis, and the phase diagram reduces to that of free electrons just as the Hamiltonian does. As the coupling constant is increased to its strong coupling limit the point $\mu_0$ decreases monotonically to the value $\mu_c=-2+2\log 2\approx -0.6137$. Let us first examine the phase diagram as $\mu$ increases along the line $B=0$. Here the ground state  has zero magnetisation and contains only bound pairs of spin-up and spin-down electrons (the $1|vw$-strings). For $\mu<-2$ the cost of having particles in the ground state is too high and the band is empty. The ground state then fills up with the $1|vw$-strings as $\mu$ is increased from $-2$ to $\mu_0$, and so is a liquid of singlet pairs. 
 When $\mu$ reaches $\mu_0$ the ground state is half-filled and remains half-filled for a range of $\mu$. Thus a gap opens for charge excitations of the ground state and phase V is insulating. 
Consider now $B>0$. As $B$ is increased spin-up electrons enter the ground state and the model becomes magnetised. This increases the value of $\mu$ that separates phases IV and V and eventually spin-polarises the model, breaking up all of the bound pairs. 

\medskip

It is possible to extend both the A- and B-models by adding the Hubbard interaction term to their Hamiltonians. This decouples the Hubbard interaction from the other terms and in particular for the B-model it allows strong onsite Coulomb repulsion of electrons to be introduced. When the extension dominates the B-model reduces to the $t$-$J$ model with coupling $J$ interpolating between $0$ and $2$ as $\uh$ changes from 0 to $\infty$. 

In general the extension destroys  the integrability of the models but it happens that there are special values of the coupling constant $\uh$ for which it does not. 
If  $\uh=0$ the extended B-model is just the Hubbard model. 
In addition  the EKS model for arbitrary $U$ can be obtained from the B-model when the coupling constant goes to infinity. Similarly 
the extended A-model    reduces to the EKS model up to the sign of the Hamiltonian after taking $\uh$ to 0. 
Our formalism can be naturally adapted to the EKS model, and we derive the simplified TBA equations and analyze the zero temperature phase diagram including the dependence on applied magnetic field, see Figure \ref{phasediags}. 

\medskip

The plan of the paper is as follows. In section 2 the models are introduced on the level of Hamiltonians and their symmetries are presented.
The Bethe equations for the Hubbard-Shastry models are discussed in section 3, and the parity invariant models are identified.  Section 4 outlines our formalism for the TBA analysis of the thermodynamic properties of the models, in particular  the string hypothesis is formulated, and the TBA equations and free energy are presented. In section 5 we investigate the TBA equations in various limits, namely the weak coupling, strong coupling, strong magnetic field, large chemical potential and infinite temperature limits. The zero temperature limit is discussed in detail in section 6 where the phase diagrams for the A- and B-models are obtained. In section 7 the integrable extensions of the A- and B-models, equivalent to the EKS model, are studied. In Conclusion we discuss 
future directions.
In the appendix A.1 our conventions, definitions and notations are summarised. The derivation of the free energy and the canonical and simplified TBA equations is outlined in appendix A.2.  The relation between repulsive and attractive models is discussed in appendix A.3 on the level of the TBA equations. The algebraic limit of the TBA equations is solved in appendix A.4. Finally in appendix A.5 the 
effect of  strong Hubbard interaction on the A-model and the extended B-model is shown to give  respectively the ferromagnetic and antiferromagnetic $t$-$J$ models.

\section{Lattice models based on Shastry's R-matrix}\la{Hams}

\subsection{The Hubbard model}

The one-dimensional Hubbard model describes spin-1/2 electrons propagating along a lattice with the following  nearest-neighbour Hamiltonian
 \bea\la{HHub}
 {\mathbf H} &=&\sum_{j=1}^L\, {\mathbf H}_{j,j+1} = \sum_{j=1}^L\, \left({\mathbf T}_{j,j+1} +4\,\uh\,{\mathbf V}^H_{j,j+1}  \right)\,,
  \eea
where $L$ is the length of the chain and 
  the kinetic (hopping) and potential terms are given by 
 \bea\la{THub}
 {\mathbf T}_{j,k} &=& -\sum_{\sigma=\uparrow,\downarrow} \big(\cd_{j,\sigma} \cm_{k,\sigma}+\cd_{k,\sigma} \cm_{j,\sigma}\big)\,,\\
 \la{UHub}
   {\mathbf V}^H_{j,k} &=&{1\ov 2} \,\big(\n_{j,\uparrow}-{1\ov 2}\big)\big(  \n_{j,\downarrow}-{1\ov 2}\big)
   						+{1\ov 2} \,\big(\n_{k,\uparrow}-{1\ov 2}\big)\big(  \n_{k,\downarrow}-{1\ov 2}\big)-{1\ov 4}\,.
  \eea
Here the canonically anticommuting fermionic operators $\cd_{j,\sigma}$  create and $\cm_{j,\sigma}$  annihilate electrons of spin $\sigma = \uparrow$ or $\sigma =\downarrow$ at the $j$-th site of the lattice.
 The operator $\n_{j,\sigma} =\cd_{j,\sigma}\cm_{j,\sigma}$ is the local particle number operator for electrons of spin $\s$ at site $j$, and in what follows we also use $\n_j=\n_{j,\uparrow}+ \n_{j,\downarrow}$. The coupling constant $\uh$ is $\uh=U/4t$, and
 the hopping parameter $t$ is set to 1. We also assume that $\uh$ is positive.
 
 \smallskip
 
 The Hamiltonian (\ref{HHub}) is very symmetric. Beyond possessing parity and translational invariance it is invariant under a spin $\su(2)$ and a charge $\su(2)$, the generators of which are given respectively by
\be\la{su2s}\begin{array}{cclcclccl}
{\mathbf  S}^+ &=& \sum_{j=1}^{L}{\mathbf  S}_j^+\,,
 & {\mathbf  S}^- &=& \sum_{j=1}^{L}{\mathbf  S}_j^-\,,
& {\mathbf  S}^z &=&\sum_{j=1}^{L} {\mathbf  S}_j^z\,,
\vspace{0.2cm}\\
{\boldsymbol \eta}^+ &=& \sum_{j=1}^{L} (-1)^j {\boldsymbol \eta}_j^+\,,
& {\boldsymbol \eta}^- &=& \sum_{j=1}^{L} (-1)^j  {\boldsymbol \eta}_j^-\,,
& {\boldsymbol \eta}^z &=& \sum_{j=1}^{L}{\boldsymbol \eta}_j^z\,.
\end{array}
\ee
where
\be\la{su2s2}\begin{array}{cclcclccl}
{\mathbf  S}_j^+ &=& \cd_{j,\uparrow} \cm_{j,\downarrow},
 & {\mathbf  S}_j^- &=& \cd_{j,\downarrow} \cm_{j,\uparrow},
& {\mathbf  S}_j^z &=&\frac{1}{2}(\n_{j,\uparrow} - \n_{j,\downarrow} )\,,
\vspace{0.2cm}
\\
 {\boldsymbol \eta}_j^+ &=& \cd_{j,\downarrow} \cd_{j,\uparrow},
& {\boldsymbol \eta}_j^- &=& \cm_{j,\uparrow} \cm_{j,\downarrow},
& {\boldsymbol \eta}_j^z &=& \frac{1}{2}(\n_{j} -1 )\,.
\end{array}
\ee
Often one considers the more general Hamiltonian
\be\la{HammuB}
{\mathbf H}-\mu\, {\mathbf N} - 2 \,B\, {\mathbf S}^z
\ee
with ${\mathbf N}=\sum_{j=1}^{L}\n_j=2{\boldsymbol \eta}^z+L$. The chemical potential $\mu$ and magnetic field $B$ break the charge and spin $\su(2)$ respectively, leaving only the two $\mathfrak{u}(1)$ subalgebras unbroken as $[{\mathbf H},{\mathbf N}]=0$ and $[{\mathbf H},{\mathbf S}^z]=0$. Note that 
\bea\la{VH}
{\mathbf V}^{\rm H} \equiv \sum_{j=1}^L\, {\mathbf V}^H_{j,j+1} = \sum_{j=1}^L\, \n_{j,\uparrow} \n_{j,\downarrow}  -{1\ov2}{\bf N}\,.
\eea
The Hubbard Hamiltonian is most commonly presented in a form that hides the charge $\su(2)$ symmetry with $\mu=-2\uh$ and $B=0$ in  \eqref{HammuB}.


There have been various attempts to extend the Hubbard model by introducing correlated hopping and adding various extra terms to the interaction of the Hamiltonian (also by introducing next-nearest-neighbour hopping but this shall not be considered here). In general one would expect these to explicitly break the symmetries of the model. It is also possible however that certain extensions realise different symmetries, e.g. \cite{EKS}-\cite{AB},  and we see this below in the A- and B-models.


\subsection{Construction} \la{Rmatconstruction}

The Hubbard-Shastry Hamiltonian  can be  obtained by using Shastry's R-matrix \cite{Shastry} and the standard techniques of integrable spin chains \cite{Faddeev, book}. 
It was found recently \cite{B06} that Shastry's R-matrix is equivalent to the $\sucex$-invariant R-matrix \cite{B05} which appears as a building block of the $\AdS$ scattering S-matrix  playing an important role in the AdS/CFT spectral problem, see \cite{AFrev,Marius} for a review.
Moreover, the $\sucex$ invariance together with the physical unitarity condition is sufficient to fix an R-matrix satisfying the Yang-Baxter equation up to a phase  \cite{B05}. 

The Hilbert space for each site of the one dimensional Hubbard model is a four dimensional graded vector space because a site can be either empty or doubly occupied, or can have a single electron of either spin. 
The fundamental representation of  $\sucex$ is also a four dimensional graded vector space $\V$, see \cite{B06} for detail. 
The Hilbert space for the model on a lattice of length $L$ is thus naturally identified with $\otimes_{j=1}^L\V_j$:
\be\la{v1bas}
 |0\rangle \leftrightarrow |e_{j,1}\rangle,\quad 
 \cd_{j,\uparrow} \cd_{j,\downarrow} |0\rangle \leftrightarrow |e_{j,2}\rangle,\quad
 \cd_{j,\uparrow} |0\rangle \leftrightarrow |e_{j,3}\rangle,\quad
 \cd_{j,\downarrow} |0\rangle \leftrightarrow |e_{j,4}\rangle \,.
 \ee
  The identification is not unique as one can perform canonical transformations on $\cm_{j,\s}\,,  \cd_{j,\s}$.

A representation of $\sucex$ depends on the values of the central elements. These can be conveniently parametrized by variables $x^\pm$ satisfying
\be\la{xpmconstr}
x^+ + \frac{1}{x^+} - x^- -\frac{1}{x^-}= 4\, i\,\uh\,.
\ee
The constraint defines a torus with $\uh$ a free parameter that characterises its elliptic modulus and that we will interpret as a coupling constant. Indeed in the setting where one obtains the Hubbard model this is precisely the coupling constant that appears in the Hubbard Hamiltonian (\ref{HHub}). 
The $\sucex$-invariant R-matrix ${\mathbf R}(x^\pm_1,x^\pm_2)$ depends on two spectral parameters which from the point of view of the $\sucex$ algebra parametrize the two 4-dim fundamental representations upon which the R-matrix acts. We restrict our attention to homogeneous spin chains, that the variables $x^\pm$ of each of the representations of $\sucex$ comprising the chain are the same at each site.
Then  $x^\pm$  play the role of extra coupling constants of the model and the density of the nearest neighbour Hubbard-Shastry  Hamiltonian is obtained in the standard way through the formula  
\be\la{scham}
{\mathbf H}_{j,j+1}\sim  i\,  {\mathbf P}_{j,j+1} \partial_1 {\mathbf R}_{j,j+1}(x^\pm_1,x^\pm)|_{x^\pm_1=x^\pm}\,,
\ee
where ${\mathbf P}_{j,j+1}$ is the permutation matrix, see appendix 
\ref{conventions} for precise definitions.

Any point $x^\pm$ on the torus can be used to obtain a homogeneous integrable spin chain model. In general however the corresponding Hamiltonian will be neither hermitian nor parity invariant. One can show by using the properties of the R-matrix that  it is hermitian if either $x^+/x^-$ or $x^+x^-$ is a phase. In the next section we examine the Bethe equations of the models and find that  there are just four distinct points on the torus, up to symmetries, that give rise to hermitian parity invariant models.\footnote{The same conclusion can be reached by using the transfer matrix and the crossing relation for the $\sucex$-invariant R-matrix.} The Hubbard model is obtained when $x^-=1/x^+$. Two of the models have not been singled out before, one at $x^-=-1/x^+$ which we call the A-model, and one at $x^-=-x^+$ which we call the B-model.\footnote{The A- and B-model Hamiltonians are particular cases of the general Hubbard-Shastry Hamiltonian derived in \cite{swad,uswad}, see eq.(12.229) of \cite{book}. In the parametrization (12.109) used in \cite{book} 
the A-model Hamiltonian (up to a unitary transformation) corresponds to  $l = i \pi/4$, and the B-model corresponds to $\mu = \pi/4$. The line $x^+x^-$ a phase corresponds to the line $Re(\mu)=0$, and the line $x^+/x^-$ a phase corresponds to the line $Re(\mu)=\pi/4$.}  
Completing the picture is the point $x^-=x^+$. Here the central extension of the algebra vanishes and one obtains the $\su(2|2)$ spin chain. It is  singular from our perspective as the coupling $\uh$ drops out.

\medskip

Given a Hamiltonian one can study either the model determined by ${\mathbf H}$ or $-{\mathbf H}$. For example instead of the Hubbard model one could consider
 \be\la{HattrHub}
 {\mathbf H}^{\rm attr. H} =\sum_{j=1}^L\, \left(-{\mathbf T}_{j,j+1} - 4\, \uh\,{\mathbf V}^H_{j,j+1}  \right)\,.
  \ee
This is the attractive Hubbard model and its physical properties are very different from the repulsive case, see e.g. \cite{book}.
For the A- and B-models, which were constructed not on physical grounds but to have rich symmetry and act on the Hubbard lattice, one has a choice for the sign of the Hamiltonian. With the hope that they capture some interesting physics the sign is chosen so that there is a cost for a doubly occupied site relative to a singly occupied site, mirroring Coulomb repulsion. Note however that although properties depend strongly on the choice of sign the two cases are closely related. This is discussed in appendix \ref{attract} at the level of the TBA equations. 

Let us comment on the sign of the kinetic term in \eqref{HattrHub}. Under the unitary transformation generated by
\bea
{\mathbf U}_1(\a) = \exp\Big( i\,\a\, \sum_{j=1}^{L} j\,( \n_{j,\uparrow} +\n_{j,\downarrow})\Big)
\eea
with $\a$ an arbitrary real parameter, the kinetic term transforms as ($k=j+1$)
\be\la{Ual}
 {\mathbf T}_{j,k} \rightarrow  {\mathbf U}_1^\dagger(\a) {\mathbf T}_{j,k}   {\mathbf U}_1(\a) = -\sum_{\sigma=\uparrow\downarrow} \big(e^{i \a}\cd_{j,\sigma} \cm_{k,\sigma}+e^{-i \a}\cd_{k,\sigma} \cm_{j,\sigma}\big)
\ee
while the interaction ${\mathbf V}^H$ is invariant. Thus the Hamiltonian \eqref{HattrHub} is unitary equivalent to 
 \be\la{HattrHub1}
 \sum_{j=1}^L\, \left({\mathbf T}_{j,j+1} -4\,\uh\,{\mathbf V}^H_{j,j+1}  \right)\,
  \ee
  under ${\mathbf U}_1(\pi)$. In momentum space $ \sum_{j=1}^L\, {\mathbf T}_{j,j+1} $ is diagonalised as $-2\sum_p \cos p\, (\n_{p,\uparrow}+\n_{p,\downarrow}) $ and the transformation ${\mathbf U}_1(\a)$ corresponds to a shift $p\rightarrow p'=p+\a$. This is why the attractive Hubbard model is conventionally written as  \eqref{HattrHub1}, as states with $p'=0$ then minimize the kinetic part of the Hamiltonian.  In general for a Hamiltonian ${\mathbf H}$ we define the Hamiltonian $\widetilde{\mathbf H}$ which has the opposite spectrum
  \be\la{opHam}
 \widetilde{\mathbf H}= -  {\mathbf U}_1^\dagger(\pi) \,{\mathbf H} \,  {\mathbf U}_1(\pi).
  \ee
  For this reason in what follows we refer to $\widetilde{\mathbf H}$ as the {\it opposite} Hamiltonian.
  
The direct calculation from the R-matrix at $x^-=1/x^+$ does not give the Hubbard Hamiltonian \eqref{HHub} but rather one must perform the unitary transformation generated by  ${\mathbf U}_1(-\pi/2)$ to obtain it, which also twists  the charge $\su(2)$ generators. To obtain the Hamiltonians for the A- and B-models (\ref{hamA}, \ref{hamB}) one must apply ${\mathbf U}_1(\pi)$ and ${\mathbf U}_1(-\pi/2)$ respectively to the corresponding results from the R-matrix calculation. These twists can be understood also at the level of the Bethe equations, see the relations \eqref{pipoints}.


\subsection{The A- and B-models}\la{HamsAB}

Both  the A- and B-models have correlated hopping kinetic terms and we will discuss these soon in turn. The interactions of the models are combinations of several common terms. In addition to the Hubbard interaction $ {\mathbf V}^H $, these are 
\bean
 {\mathbf V}_{j,k}^{CC} & = &\,{\boldsymbol \eta}_j^z\, {\boldsymbol \eta}_k^z-\frac{1}{4}\ =\ \frac{1}{4}\left(\n_{j}-1)( \n_{k}-1 \right)-\frac{1}{4}\ \,,\\
 {\mathbf V}_{j,k}^{SS} &=& \frac{1}{2}(	{\mathbf  S}_j^+\, {\mathbf  S}_k^-+{\mathbf  S}_j^- \,{\mathbf  S}_k^+ )
 	+ {\mathbf  S}_j^z\, {\mathbf  S}_k^z\\
& = & \frac{1}{2}(\cd_{j,\uparrow} \cm_{j,\downarrow} \cd_{k,\downarrow} \cm_{k,\uparrow}+\cd_{j,\downarrow} \cm_{j,\uparrow} \cd_{k,\uparrow} \cm_{k,\downarrow} )
 	+\frac{1}{4} \left(\n_{j,\uparrow} - \n_{j,\downarrow} \right)\left(\n_{k,\uparrow} - \n_{k,\downarrow} \right)\,,\\
  {\mathbf V}_{j,k}^{PH} & = &\frac{1}{2}( {\boldsymbol \eta}_j^+\, {\boldsymbol \eta}_k^- +{\boldsymbol \eta}_j^-\, {\boldsymbol \eta}_k^+ )\ =\  \frac{1}{2}(\cd_{j,\uparrow} \cd_{j,\downarrow} \cm_{k,\downarrow} \cm_{k,\uparrow} +\cd_{k,\uparrow} \cd_{k,\downarrow}   \cm_{j,\downarrow}\cm_{j,\uparrow})\,.
\eean
The charge-charge interaction $ {\mathbf V}^{CC}$ has the interpretation of nearest neighbour Coulomb repulsion. Physically one should think of a site as having charge +1, and so a site with one electron on it has neutral charge, a site with electron density less than one is positively charged and a site with electron density greater than one is negatively charged. The spin-spin interaction $ {\mathbf V}^{SS} $ is the familiar spin-exchange term of the Heisenberg XXX spin chain. Finally the pair hopping interaction ${\mathbf V}^{PH}$ relates to the simultaneous hopping of two electrons from one site to a neighbouring site.

Now we discuss the A- and B-models in turn. The presentation of the Hamiltonians is much cleaner if  the coupling constant $\uh$ is re-expressed  through 
$$\uh= \sinh \nu\,.$$

\subsubsection*{A-model}

The model at $x^+x^-=-1$ has the Hamiltonian
 \bea\nonumber
  {\mathbf H}^{\rm A} &=& \sum_{j=1}^L\, \Big( {\mathbf T}_{j,j+1}^{\rm A} +\frac{2 \cosh 2\nu}{\cosh\nu}\, {\mathbf V}^H_{j,j+1}
  		+{2\ov \cosh \nu}\,\big(  {\mathbf V}_{j,j+1}^{CC} -{\mathbf V}_{j,j+1}^{SS} + {\mathbf V}_{j,j+1}^{PH} \big) \Big)\,,~~~~~~\\ \la{hamA}
  {\mathbf T}_{j,k}^{\rm A} &=&  - \sum_{\sigma} \Big[\big( \cd_{j,\sigma} \cm_{k,\sigma}+ \cd_{k,\sigma} \cm_{j,\sigma}\big)\big( 1-\n_{j,-\sigma} - \n_{k,-\sigma} \big) \\
 & &\qquad\qquad- i \tanh\nu\big( \cd_{j,\sigma} \cm_{k,\sigma}- \cd_{k,\sigma} \cm_{j,\sigma}\big)\big(\n_{j,-\sigma} - \n_{k,-\sigma}\big)^2{\Big]}.
 \notag
  \eea
It is easy to check that it is hermitian  but its parity invariance is more subtle. 
Note that the transformation
  \bea\la{partr}
  \cd_{j,\sigma}\rightarrow \cd_{L-j+1,\sigma}\, , \qquad \cm_{j,\sigma}\rightarrow \cm_{L-j+1,\sigma}\, ,
  \eea
  is not a symmetry but is instead equivalent to a replacement $\nu\rightarrow -\nu$ in the Hamiltonian. This however is just a unitary transformation. More generally, the Hamiltonian \eqref{hamA} is unitary equivalent to the following one
 \bea\notag
   {\mathbf H}^{\rm A}(\a) &\equiv& {\mathbf U}_2^\dagger(\a)\,  {\mathbf H}^{\rm A} \,{\mathbf U}_2(\a)\,,\quad {\mathbf U}_2(\a)\equiv\exp\left(i\,\a\,\sum_{j=1}^L \n_{j,\uparrow} \n_{j,\downarrow}\right)\,, \notag\\\notag
  {\mathbf H}^{\rm A}(\a) &=& \sum_{j=1}^L\, \Big( {\mathbf T}_{j,j+1}^{\rm A}(\a) +\frac{2 \cosh 2\nu}{\cosh\nu}\, {\mathbf V}^H_{j,j+1}
  		+{2\ov \cosh \nu}\,\big(  {\mathbf V}_{j,j+1}^{CC} -{\mathbf V}_{j,j+1}^{SS} + {\mathbf V}_{j,j+1}^{PH} \big) \Big)\,,\\
    {\mathbf T}_{j,k}^{\rm A}(\a) &=&   - \sum_{\sigma} \Big[\big( \cd_{j,\sigma} \cm_{k,\sigma}+ \cd_{k,\sigma} \cm_{j,\sigma}\big)\big( 1-\n_{j,-\sigma} - \n_{k,-\sigma} -\sin\a \tanh\nu\,(\n_{j,-\sigma} - \n_{k,-\sigma})\big) \notag \\
 & &\qquad\qquad- i\cos\a \tanh\nu\big( \cd_{j,\sigma} \cm_{k,\sigma}- \cd_{k,\sigma} \cm_{j,\sigma}\big)\big(\n_{j,-\sigma} - \n_{k,-\sigma}\big)^2{\Big]}
\,,
 \la{HA2}
  \eea
  where $\a$ is an arbitrary real parameter. If one chooses $\a=\pi$ one gets 
  $$
   {\mathbf H}^{\rm A}(\pi) = {\mathbf H}^{\rm A} \ \ {\rm with}\ \ \nu\to-\nu\,, 
  $$
  and so the parity transformation \eqref{partr} should be supplemented with the change of basis generated by $ {\mathbf U}_2(\pi)$.
 Note also that with the choice $\a=\pm\pi/2$ one removes the imaginary term from the Hamiltonian.

The kinetic part of the Hamiltonian is of a complicated correlated hopping type. It is instructive to see how it acts in various cases. For an electron of spin $\sigma$ to hop between two sites $j$ and $k$ they must not contain other electrons of that spin but  they can contain electrons of the opposite spin. We consider individually the four possibilities as  illustrated in Figure \ref{hopping} for a mobile spin-up electron. The kinetic density $ {\mathbf T}_{j,k}^{\rm A}(\a) $ acting on each of the cases takes the form
\begin{figure}[t]
\includegraphics[width=\linewidth]{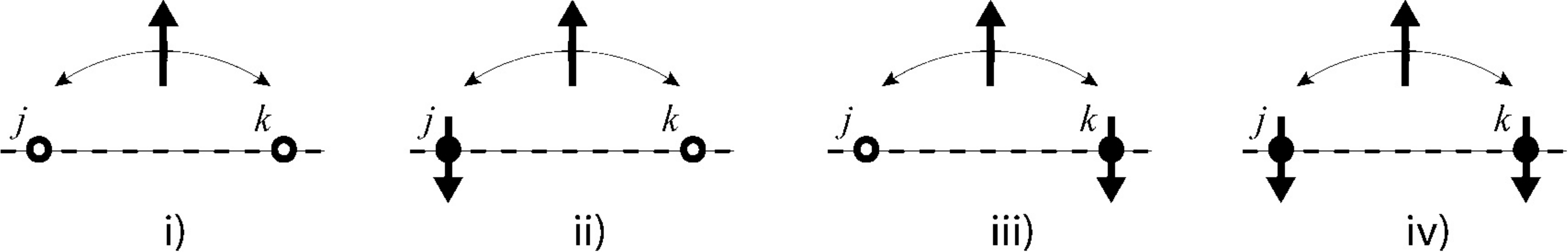}
\caption{\small The four cases of an electron hopping between sites $j$ and $k$: i) no electrons of the opposite spin on either site, ii) an electron of the opposite spin on site $j$ only, iii) an electron of the opposite spin on site $k$ only, iv) electrons of the opposite spin on both sites.}
  \la{hopping}
\end{figure}
\begin{enumerate}[label=\roman{*})]
\item[] i) $\quad$    $-  \cd_{j,\sigma} \cm_{k,\sigma}-\, \cd_{k,\sigma} \cm_{j,\sigma}$,
\item[] ii) $\quad$  $ - i e^{-i\a} \tanh\nu\,  \cd_{j,\sigma} \cm_{k,\sigma}+ i e^{i \a} \tanh{\nu} \,\cd_{k,\sigma} \cm_{j,\sigma}$,
\item[] iii) $\quad$  $ - i e^{i\a} \tanh\nu \, \cd_{j,\sigma} \cm_{k,\sigma}+ i e^{-i \a} \tanh{\nu}\, \cd_{k,\sigma} \cm_{j,\sigma}$,
\item[] iv) $\quad$    $  \cd_{j,\sigma} \cm_{k,\sigma}+\, \cd_{k,\sigma} \cm_{j,\sigma}$.
\end{enumerate}
Note the different sign between cases i) and iv),  while in the case of free electrons all the cases would be the same as case i). This reflects the strong electron correlation in the A-model which cannot be reduced to free electrons for any coupling.

Like all the models we consider, the A-model inherits spin  and charge $\su(2)$ symmetries from the $\sucex$ structure. The spin $\su(2)$ generators take the same form as those for the Hubbard model \eqref{su2s} whereas the charge $\su(2)$ generators take a untwisted  form.

The weak coupling $\nu\rightarrow0$ limit of the A-model is not described by free electrons but rather the Hamiltonian reduces to the ${\su(2|2)}$ spin chain. This has received much study due to an interesting generalisation found in \cite{EKS}. A key observation is that  ${\mathbf V}^{\rm H}$  commutes with ${\mathbf H}^{\rm A}\big|_{\nu=0}$ and so one can extend the model, as will be discussed in section \ref{exmods}. 
 In the strong coupling  $\nu\rightarrow\infty$ limit only the term $4\,\uh {\mathbf V}^{\rm H}$ in the Hamiltonian survives and it thus shares the strict strong coupling limit of the Hubbard model.  In section \ref{strongcoupling} we show that at half-filling and strong coupling the model behaves as a ferromagnetic XXX spin chain. 
 The  strong coupling limit of the less than half-filled model is analysed in appendix \ref{tJB} and is found to be a ferromagnetic   {\it t-J}  model.

\subsubsection*{B-model}
The model at $x^+/x^-=-1$ has the Hamiltonian
 \bea\la{hamB}
  {\mathbf H}^{\rm B} &=& \sum_{j=1}^L\, \Big( {\mathbf T}_{j,j+1}^{\rm B} +2 \tanh\nu\,\big( {\mathbf V}_{j,j+1}^{H}- {\mathbf V}_{j,j+1}^{CC}+{\mathbf V}_{j,j+1}^{SS}
  		+{\mathbf V}_{j,j+1}^{PH} \big) \Big)\,,\\
  {\mathbf T}_{j,k}^{\rm B} &=& - \sum_{\sigma} \big(\cd_{j,\sigma} \cm_{k,\sigma}+\cd_{k,\sigma} \cm_{j,\sigma}\big)
  \Big(1- \big(1-{1\ov \cosh \nu}\big) \big(\n_{j,-\sigma} - \n_{k,-\sigma}  \big)^2\Big)\notag.
  \eea
Let us begin the analysis by examining the correlated hopping kinetic term for the four cases illustrated in Figure \ref{hopping}. The model is clearly parity invariant and so cases  ii) and iii) are identical. The kinetic density $ {\mathbf T}_{j,k}^{\rm B}(\a) $ acting on each of the cases takes the form
\begin{itemize}
\item[] i), iv) $\quad$   $-  \cd_{j,\sigma} \cm_{k,\sigma}-\, \cd_{k,\sigma} \cm_{j,\sigma}$,
\item[] ii), iii) $\,\,\,\,$ $ - {1\ov \cosh \nu}(\cd_{j,\sigma} \cm_{k,\sigma}+\, \cd_{k,\sigma} \cm_{j,\sigma})$.
\end{itemize}
In cases i) and iv) hopping is of the standard form, while in cases ii) and iii) its contribution is proportional to $\mbox{sech}\,\nu$ and decreases to zero in the strong coupling limit.
 Both cases i) and iv) preserve the number of doubly occupied sites, while for hopping in cases ii) and iii) the number will be either increased or decreased by 1.

The B-model describes free electrons in the weak coupling $\nu\rightarrow 0$ limit. In the strong coupling $\nu\rightarrow\infty$  limit note that $[{\mathbf H}^{\rm B}\,,{\mathbf V}^{\rm H}]=0$, as for the weak coupling limit of the A-model. This allows one to decouple the coefficient of the Hubbard interaction from the other terms while preserving the integrability, see section \ref{exmods}.

\smallskip

The B-model possesses a rich symmetry structure. Its spin and charge $\su(2)$  generators take the same form as for the Hubbard model \eqref{su2s}. Moreover the model inherits a hidden $\su(2|2)$ fermionic symmetry for arbitrary coupling constant $\nu$ from Shastry's R-matrix. There are eight supercharges, four of which ${\mathbf Q}^\dagger_{0,\sigma}$, ${\mathbf Q}_{0,\sigma}$  create/annihilate an electron with momentum $p=0$, and the other four ${\mathbf Q}^\dagger_{\pi,\sigma}$, ${\mathbf Q}_{\pi,\sigma}$  create/annihilate an electron with momentum $p=\pi$.
Explicitly they are
{\small\bea\la{QB}
{\mathbf Q}_{\a, \sigma}&=&\sum_{j=1}^{L} e^{i\, \a \,j}(\gamma_{\a}\, \cm_{j,\sigma}- \beta\, \n_{j,-\sigma}\cm_{j,\sigma})\,,\quad {\mathbf Q}^\dagger_{\a, \sigma}=({\mathbf Q}_{\a, \sigma})^\dagger\,,\quad \a=0,\pi\,,~~~~\\\nonumber
&\mbox{with}&\gamma_{\a}={1\ov2} \left(\beta+{e^{i\, \a} \ov\beta}\right)\mbox{ and }\beta = \sqrt{\tanh{\nu \ov 2}}\,,
\eea}and they satisfy the $\su(2|2)$ algebra with central charge $\coth\nu$:
{\small \bean
 \{{\mathbf Q}_{0,\downarrow},{\mathbf Q}^\dagger_{0,\uparrow}\}&=&
-\, \{{\mathbf Q}_{\pi,\downarrow},{\mathbf Q}^\dagger_{\pi,\uparrow}\}\,=\, {\mathbf S}^+ \,,\quad
 \{{\mathbf Q}_{0,\uparrow},{\mathbf Q}^\dagger_{0,\downarrow}\}\,=\, 
-\,  \{{\mathbf Q}_{\pi,\uparrow},{\mathbf Q}^\dagger_{\pi,\downarrow}\}\,=\, {\mathbf S}^-\,,\\
\{{\mathbf Q}^\dagger_{0,\downarrow},{\mathbf Q}^\dagger_{\pi,\uparrow}\}&=& 
-\,\{{\mathbf Q}^\dagger_{0,\uparrow},{\mathbf Q}^\dagger_{\pi,\downarrow}\}\,=\, {\boldsymbol \eta}^+ \, ,\quad\,
\{{\mathbf Q}_{0,\downarrow},{\mathbf Q}_{\pi,\uparrow}\}\,=\,
-\,\{{\mathbf Q}_{0,\uparrow},{\mathbf Q}_{\pi,\downarrow}\} \,=\, {\boldsymbol \eta}^-\,,\\
  \{{\mathbf Q}_{0,\uparrow},{\mathbf Q}^\dagger_{0,\uparrow}\}&=& {\mathbf S}^z-{\boldsymbol \eta}^z+\coth\nu \,, \qquad
  \quad  \{{\mathbf Q}_{0,\downarrow},{\mathbf Q}^\dagger_{0,\downarrow}\}\,=\,-\, {\mathbf S}^z-{\boldsymbol \eta}^z+\coth\nu  \,,\\ 
    \{{\mathbf Q}_{\pi,\uparrow},{\mathbf Q}^\dagger_{\pi,\uparrow}\}&=& -\,{\mathbf S}^z+{\boldsymbol \eta}^z+\coth\nu \,,   \qquad\,
    \{{\mathbf Q}_{\pi,\downarrow},{\mathbf Q}^\dagger_{\pi,\downarrow}\}\,=\, {\mathbf S}^z+{\boldsymbol \eta}^z+\coth\nu\,,
\eean 
}and all other anti-commutators vanish.
The symmetry is hidden in that these generators do not commute with the Hamiltonian but rather
{\small\bean
[{\mathbf H^{\rm B}},{\mathbf Q}] = h_Q \,{\mathbf Q}\,,\quad [{\mathbf N},{\mathbf Q}] = n_Q \,{\mathbf Q}\,,\quad[{\mathbf S}^z,{\mathbf Q}] = s_Q \,{\mathbf Q}\,,
\eean}with $h_Q$, $n_Q$ and $s_Q$ given in Table \ref{hnsq1}. 
\begin{table}[t]
\begin{center}
\begin{tabular}{c|cccccccc}
 & ${\mathbf Q}_{0, \uparrow}$ & ${\mathbf Q}_{0, \downarrow}$& ${\mathbf Q}_{\pi, \uparrow}$& ${\mathbf Q}_{\pi, \downarrow}$& ${\mathbf Q}^\dagger_{0, \uparrow}$ & ${\mathbf Q}^\dagger_{0, \downarrow}$& ${\mathbf Q}^\dagger_{\pi, \uparrow}$& ${\mathbf Q}^\dagger_{\pi, \downarrow}$\\
\hline
$h_Q$ & 2 & 2 & -2 & -2 & -2 & -2 & 2 & 2 \\
$n_Q$ & -1 & -1 & -1 & -1 & 1 & 1 & 1 & 1 \\
$s_Q$ & -1/2 & 1/2 & -1/2 & 1/2 & 1/2 & -1/2 & 1/2 & -1/2 \\
\end{tabular}
\end{center}
\caption{\small Commutators of supercharges of the B-model with $\mathbf{H}$, $\mathbf{N}$ and $\mathbf{S}^z$.}
\la{hnsq1}
\end{table}
Thus the symmetry does not lead to an extra degeneracy in the spectrum of the model, but it does imply exact relationships between the energies of eigenstates. For example if $|\psi\rangle$ is an eigenstate  with energy $\E_{\psi}$, then the energy of the state {\small${\mathbf Q}|\psi\rangle$} is given through  {\small${\mathbf H}^{\rm B} {\mathbf Q}|\psi\rangle=(\E_{\psi}+h_{ Q}){\mathbf Q}|\psi\rangle$}. In particular it follows that the ground state of {\small${\mathbf H}^{\rm B}$} is in the kernel of  {\small${\mathbf Q}^\dagger_{0, \uparrow}$,  ${\mathbf Q}^\dagger_{0, \downarrow}$, ${\mathbf Q}_{\pi,\uparrow}$} and {\small${\mathbf Q}_{\pi,\downarrow}$}.

Let us conclude with two remarks. First, the above fermionic symmetry can be generalised naturally to any bipartite lattice in any dimension. This raises the prospect of being able to extract some exact information for such lattices. Finally, although the symmetry is hidden for {\small${\mathbf H^{\rm B}}$}, it can be realised in part for the Hamiltonian {\small${\mathbf H^{\rm B}}-\mu\, {\mathbf N} - 2 \,B\, {\mathbf S}^z$} for appropriately  tuned $\mu$ and $B$. To this end it is useful to consider not just the above supercharges  {\small${\mathbf Q}$} but also products of them with {\small$ {\mathbf  S}^+$, $ {\mathbf  S}^-$, $ {\boldsymbol \eta}^+$} and {\small$ {\boldsymbol \eta}^-$}.

\subsection{Extended models}\la{exmods}

The Hubbard interaction term preserves both spin and charge $\su(2)$ symmetries. It is physically reasonable to 
add this term with an arbitrary coupling $\gh$ to the Hamiltonians of the A- and B-models
and  the opposite models  because it allows strong onsite Coulomb repulsion of electrons to be introduced.  In particular 
the $\gh\to\infty$ limit is studied  in appendix \ref{tJB} where it is shown that the B-model reduces to the $t$-$J$ model with the coupling interpolating between $J=0$ and $J=2$ as $\uh$ changes from 0 to $\infty$.

In general the extension destroys  the integrability of the models but there are special values of $\uh$ for which it does not. 
In this section we present integrable extensions of the weak coupling limit of the A-model and the strong coupling limit of the B-model. We borrow an idea from \cite{EKS} and the model ${\mathbf H}^{\rm {\tilde A}_0}$ below is indeed the one they studied. The key observation is that if for a model  $[{\mathbf H}\,,{\mathbf V}^{\rm H}]=0$ then ${\mathbf H}$ and ${\mathbf V}^{\rm H}$ can be diagonalised simultaneously and it is natural to add a term $4\,\gh \,{\mathbf V}^{\rm H}$ to the Hamiltonian, where $\gh$ is a new coupling constant. This allows one to decouple the term capturing onsite Coulomb repulsion from the other terms in  the Hamiltonian. Since the number of doubly occupied sites is conserved if $[{\mathbf H},{\mathbf N}]=0$, then $\gh$ can be thought of as a chemical potential for doubly occupied sites. Furthermore if the original model is integrable and ${\mathbf V}^{\rm H}$ commutes with the transfer matrix then the resulting model is integrable too. This is the case for the limits of the A- and B-models that we consider here.

The extended Hamiltonians for the weak coupling limit of the A-model and its opposite model (see equation \eqref{opHam}) are
\bea\la{HamA0}
 {\mathbf H}^{\rm A_0}(\gh) &=& \sum_{j=1}^L\, \Big( {\mathbf T}_{j,j+1}^{\rm A_0} 
 		+2\big((2\,\gh+1)\, {\mathbf V}^H_{j,j+1}+{\mathbf V}_{j,j+1}^{CC}-{\mathbf V}_{j,j+1}^{SS}+{\mathbf V}_{j,j+1}^{PH}\big)  \Big)\,, \\\la{opHamA0}
  {\mathbf H}^{\rm \tilde A_0}(\gh) &=& \sum_{j=1}^L\, \Big( {\mathbf T}_{j,j+1}^{\rm A_0}  
  		+2\big((2\,\gh-1)\, {\mathbf V}^H_{j,j+1}-{\mathbf V}_{j,j+1}^{CC}+{\mathbf V}_{j,j+1}^{SS}-{\mathbf V}_{j,j+1}^{PH}\big)  \Big)\,,\\
  {\mathbf T}_{j,k}^{\rm A_0} &=&  - \sum_{\sigma} \big(\cd_{j,\sigma} \cm_{k,\sigma}+\, \cd_{k,\sigma} \cm_{j,\sigma}\big) \big( 1 -\n_{j,-\sigma} - \n_{k,-\sigma}\big)\,. \notag
  \eea
 The Hamiltonian ${\mathbf H}^{\rm \tilde A_0}(\gh)$ 
is the one for the EKS model \cite{EKS} (with $\gh= U/4$), and eq.\eqref{opHamA0} shows that for $\gh<1/2$ it should be thought of as a model with attractive on-site Coulomb interaction.
The extended Hamiltonians  for the strong coupling limit of the B-model and its opposite model  are
\bea\la{HamBinf}
  {\mathbf H}^{\rm B_\infty}(\gh) &=& \sum_{j=1}^L\, \Big( {\mathbf T}_{j,j+1}^{\rm B_\infty}  
  		+2\big((2\,\gh+1)\, {\mathbf V}^H_{j,j+1}-{\mathbf V}_{j,j+1}^{CC}+{\mathbf V}_{j,j+1}^{SS}+{\mathbf V}_{j,j+1}^{PH} \big) \Big)\,,\\  \la{opHamBinf}
 {\mathbf H}^{\rm \tilde B_\infty}(\gh) &=& \sum_{j=1}^L\, \Big( {\mathbf T}_{j,j+1}^{\rm B_\infty}  
  		+2\big((2\,\gh-1)\, {\mathbf V}^H_{j,j+1}+{\mathbf V}_{j,j+1}^{CC}-{\mathbf V}_{j,j+1}^{SS}-{\mathbf V}_{j,j+1}^{PH} \big) \Big)\,,\\
  {\mathbf T}_{j,k}^{\rm B_\infty} &=& - \sum_{\sigma} \big(\cd_{j,\sigma} \cm_{k,\sigma}+\cd_{k,\sigma} \cm_{j,\sigma}\big)
  \big(1-  (\n_{j,-\sigma} - \n_{k,-\sigma})^2\big)\notag.
  \eea
  
  Similar to the B-model, both extended models have hidden fermionic symmetry. 
 Since the extended A-model with $\gh=0$ is the $\su(2|2)$ spin chain its Hamiltonian commutes with the supercharges \cite{EKS} which create and annihilate electrons on singly occupied and doubly occupied sites
\bea\la{QA2}
{\mathbf Q}_{s,\sigma}=\sum_{j=1}^{L} ( \cm_{j,\sigma}- \n_{j,-\sigma}\cm_{j,\sigma})\,,\quad {\mathbf Q}_{d, \sigma}=\sum_{j=1}^{L} \, \n_{j,-\sigma}\cm_{j,\sigma}\,,
\eea
and their hermitian conjugates. Noting that 
\bea
[ {\mathbf V}^{\rm H}\,, {\mathbf Q}_{s, \sigma}] = {1\ov 2}{\mathbf Q}_{s, \sigma}\,,\quad [ {\mathbf V}^{\rm H}\,, {\mathbf Q}_{d, \sigma}] = -{1\ov 2}{\mathbf Q}_{d, \sigma}\,,
\eea
one finds
\bea
[  {\mathbf H}^{\rm A_0}(\gh) \,, {\mathbf Q}_{s, \sigma}] = 2\gh\, {\mathbf Q}_{s, \sigma}\,,\quad [  {\mathbf H}^{\rm A_0}(\gh) \,, {\mathbf Q}_{d, \sigma}] = -2\gh\, {\mathbf Q}_{d, \sigma}\,.
\eea
Taking the limit $\uh\to\infty$ for the supercharges \eqref{QB} of the B-model one finds similarly
 \bea\la{QB2}
{\mathbf Q}_{0,\sigma}=\sum_{j=1}^{L} ( \cm_{j,\sigma}- \n_{j,-\sigma}\cm_{j,\sigma})\,,\quad {\mathbf Q}_{\pi, \sigma}=\sum_{j=1}^{L} \, (-1)^j\n_{j,-\sigma}\cm_{j,\sigma}\,,
\eea
and 
\bea
[  {\mathbf H}^{\rm B_\infty}(\gh) \,, {\mathbf Q}_{0, \sigma}] = 2(\gh+1)\, {\mathbf Q}_{0, \sigma}\,,\quad [  {\mathbf H}^{\rm B_\infty}(\gh) \,, {\mathbf Q}_{\pi, \sigma}] = -2(\gh+1)\, {\mathbf Q}_{\pi, \sigma}\,.
\eea  
  
The Hamiltonians and symmetries of these models obviously look  different but, as we will see in section \ref{TBAext}, for even $L$ they have the same Bethe equations. Analyzing their dispersion relations one finds that 
the Hamiltonians  ${\mathbf H}^{\rm B_\infty}(\gh)$ and  ${\mathbf H}^{\rm\tilde A_0}(\gh+1)$ have the same spectrum and therefore they should be unitary equivalent. 
The corresponding unitary transformation is given by 
\be\la{ut}
{\mathbf U}\equiv\exp\left(i\,\pi\,\sum_{j=1}^L\, j\,\n_{j,\uparrow} \n_{j,\downarrow}\right)\,,
\ee 
and it shifts the momenta of doubly occupied sites by $\pi$ while leaving the momenta of singly occupied sites unchanged. 
Correspondingly it untwists  
the supercharges ${\mathbf Q}_{\pi, \sigma}$ of the extended B-model but leave unchanged  ${\mathbf Q}_{0, \sigma}$.

\section{Bethe equations}\la{BEqns}
In the previous section a family of models with Hamiltonian (\ref{scham}) were introduced. These were constructed from the $\sucex$ invariant Shastry R-matrix using the quantum inverse scattering method. The models have been diagonalised for arbitrary $x^\pm$ using the algebraic Bethe ansatz  \cite{MaRa,MaMe,Marius}.  For the purpose of identifying the momentum and dispersion relation, the relevant part of the eigenvalue of the transfer matrix ${\mathbf t}$ for a state on a periodic one-dimensional chain of length $L$ with $N$ charge excitations over the pseudo-vacuum $(|0\rangle)^L$ is given by, see e.g.  \cite{B06,ALST}
\bea\la{teval}
t=\prod_{k=1}^{N}e^{i \phi}\frac{y_k-x^-}{y_k-x^+}\sqrt{\frac{x^+}{x^-}},
\eea
where $\phi\in{\mathbb R}$ is an arbitrary twist which should satisfy the condition $\exp(i\phi L)=1$ for a periodic spin chain. The $y_k$ are the Bethe roots for charge excitations.
The momentum is then defined through
\bea\la{pk}
e^{ip_k} = e^{i \phi} \frac{y_k-x^-}{y_k-x^+}\sqrt{\frac{x^+}{x^-}}\,,\quad -\pi\le p_k\le\pi\,,
\eea
as the  transfer matrix is the shift operator and so its eigenvalue is naturally identified with $e^{i \sum_k p_k}$.
The roots $y_k$ are determined from the Bethe equations
\be\la{BE1}
\begin{aligned}
e^{i p_k L} =& \,
\prod_{j=1}^{M}\frac{v_k-w_j+i\,\uh}{v_k-w_j-i\,\uh}\,, & k=1,\ldots,N \, \le L\,,  \\
\prod_{j=1}^{N}\frac{w_k-v_j+i\,\uh}{w_k-v_j-i\,\uh}
=& \prod_{j=1,j\neq
k}^{M}\frac{w_k-w_i+2i\,\uh}{w_k-w_j-2i\,\uh}\,,& k=1,\ldots,M \, \le N/2\,,
\end{aligned}
\ee
where $w$ are the Bethe roots for spin excitations,\footnote{Our notations are related to the ones adopted in \cite{book} as $p_j \leftrightarrow k_j$, $w_j \leftrightarrow \Lambda_j$.} and the $v$ are related to the charge Bethe roots via  $v=\frac{1}{2}(y+1/y)$.

As stated in the previous section the models will only be hermitian if either $x^+/x^-$ or $x^+x^-$ is a phase. The requirement of parity invariance is more stringent.\footnote{We thank Niklas Beisert for a valuable discussion of this issue.} At the level of the Bethe equations it amounts to a symmetry under $p\rightarrow -p$. 
There are two distinct ways this can be realised
\bea\begin{aligned}
p\rightarrow -p,\quad y\rightarrow -y, \quad w\rightarrow -w\,,~~~~\la{transf1}\\
p\rightarrow -p,\quad y\rightarrow -\frac{1}{y}, \quad w\rightarrow -w\,.~~~~
\end{aligned}\eea
However $p$ and $y$ are related through (\ref{pk}) and so together with equation (\ref{xpmconstr}) these transformations strongly constrain $x^\pm$ and $\phi$. One finds that there are four possible solutions leading to four distinct parity invariant models, each of which satisfy the requirement to be hermitian\footnote{The fourth solution $x^+=x^-=\infty$ corresponds to the $\su(2|2)$ spin chain but this point is highly degenerate and to get the correct Bethe equations for the model from \eqref{BE1} one should send the roots $v_k$ and $w_k$ to infinity in a proper way.  Technically it is similar to the way we get the Bethe equations for the $\su(2|2)$ spin chain from the weak coupling limit of the A-model in section \ref{TBAext}.}
\begin{equation}\la{pipoints}\begin{array}{cclcl}
({\rm H})&\quad&x^+=1/x^-=\infty, &\quad &\phi=\frac{2n-1}{2} \pi\,,\\
({\rm A})&\quad&x^+=-1/x^-=i(\uh+\sqrt{1+\uh^2}), &\quad &\phi=n \pi\,,\\
({\rm B})&\quad&x^+=-x^-=i(\uh+\sqrt{1+\uh^2}), &\quad &\phi=\frac{2n-1}{2} \pi\,,\quad n=0,1\,.\\
\end{array}\end{equation}
In what follows we choose $\phi_{\rm H}=-{\pi\ov 2}$, $\phi_{\rm A}=\pi$ and $\phi_{\rm B}=-{\pi\ov 2}$ which lead to the minus sign in front of the $\cos p$ term in the dispersion relations, see \eqref{disprels}. 

Next we wish to obtain the dispersion relation of these models. To proceed one observes how the Hamiltonian acts on one particle states. The result can be obtained from the explicit forms of the A- and B-model Hamiltonians (\ref{hamA}, \ref{hamB}), but it is of interest to obtain it in the present context also. The transfer matrix is related to the Hamiltonian through ${\mathbf H}\sim i \, {\mathbf t}^{-1} \partial {\mathbf t}$, and so its eigenvalue for a single particle, the dispersion relation, follows straightforwardly from the transfer matrix eigenvalue (\ref{teval}) with $N=1$. For the four models, using the same normalisation as for the Hamiltonians of the previous section, one finds
\begin{equation}\la{disprels}\begin{array}{ccl}
({\rm H})&\quad &\E(p) = - 2 \cos (p)-2\,\uh.\\
({\rm A})&\quad &\E(p) = - 2 \cos (p)-2\sqrt{1+\uh^2},\\
({\rm B})&\quad&\E(p) = - 2 \cos  (p).\\
\end{array}\end{equation}
This clearly agrees with what one expects from the explicit forms of the Hamiltonians.

Let us conclude by remarking how the transformation between a model and its opposite, given by equation \eqref{opHam}, can be understood in the present context. Shifting  $p\rightarrow p+\pi$, which is equivalent to choosing the other value for $n$ in \eqref{pipoints}, and changing the overall sign of the dispersion relation \eqref{disprels}, provides the dispersion relation for the opposite model.

\section{Thermodynamic limit}

Let us turn our attention to the physical properties of the models introduced above. We will examine the thermodynamic limit, $L\rightarrow\infty$, in the grand canonical ensemble. The equilibrium thermodynamic properties follow from the minimization of the free energy
\be\la{deffreng}
f(\mu,B,T)=-\frac{T}{L} \log \left( \tr \left[\exp\left(-\frac{{\mathbf H}-\mu\, {\mathbf N} - 2 \,B\, {\mathbf S}^z}{T}\right)\right]\right).
\ee
Here $\mu$ is the chemical potential, $B$ is the  magnetic field, and $T$ is the temperature. 

The magnetization $m$ per site, particle density $n_c$, specific heat capacity $c$, and the spin $ \chi_s$ and charge $ \chi_c$ susceptibilities are calculated in the standard way
\bea\la{physprops}
m &=& -\frac{1}{2} \frac{\partial f}{\partial B},\quad n_c = - \frac{\partial f}{\partial \mu},\quad 
c=-T \frac{\partial^2 f}{\partial\, T^2},\quad \chi_s = \frac{\partial\, m}{\partial\,B},\quad \chi_c = \frac{\partial\, n_c}{\partial\, \mu}.
\eea
The  free energy obeys the following symmetries
\be\la{frengsym}
f(\mu,-B,T)=f(\mu,B,T)\,,\quad
f(-\mu,B,T)=f(\mu,B,T)+2\mu\,.
\ee
The first follows from the invariance of the Hamiltonians of the models with respect to spin reversal. To see the second note that  ${\mathbf N}\rightarrow 2 L -{\mathbf N}$ under the particle-hole transformation
$\cd_{j,\sigma}\leftrightarrow (-1)^j \cm_{j,-\sigma}$
for the Hubbard and B-models, 
and $\cd_{j,\sigma}\leftrightarrow  \cm_{j,-\sigma}$ for the A-model, which is also a symmetry of the Hamiltonians.

\subsection{String hypothesis}\la{stringhyp}

The free energy is calculated in the thermodynamic limit using the Bethe ansatz and making a string hypothesis \cite{Takahashi} for the behaviour of the roots in this limit. 

To describe the string hypothesis and TBA equations it is convenient  to introduce  the following functions
\bea
x_{\rm A}(v)=v+v\sqrt{1-\frac{1}{v^2}}\,,\quad |x_{\rm A}(v)|\ge 1\,,\  v\in {\mathbf C}\,,
\eea
with a cut $\I^{\rm A}=(-1,1)$, and 
\bea
x_{\rm B}(v)=v+i\sqrt{1-v^2}\,, \quad {\rm Im}(x_{\rm B}(v))\ge 0\,,\  v\in {\mathbf C}\,,
\eea
with a cut $\I^{\rm B}=(-\infty,-1)\cup (1,\infty)$. Let us also define $\I^{\rm A, B}_+$ and $\I^{\rm A, B}_-$ to be respectively the upper and lower edges of $\I^{\rm A, B}$. For values of $v$ on the cuts we define $x(v)=x(v+i0)$.
Both functions solve the equation
$$
x(v)+\frac{1}{x(v)}=2v\,,
$$
and the parameters $x^\pm$ of the A-  and  B-models  can be uniformly written as $x^+ = x(i\, \uh)$, 
$x^-= 1/x(-i\, \uh)$
 where $x = x_{\rm A}$ and  $x = x_{\rm B}$ for the A- and the B-model, respectively.\footnote{
The parameters $x^\pm$ of the 
two families of Hermitian (but in general not parity-invariant) Hubbard-Shastry models can be written as $x^+ = x(\lam+i\, \uh)$,  $x^-= 1/x(\lam-i\, \uh)$  where $\lam$ is an arbitrary real number. 
For the models with $x^+x^-$ being a phase $x= x_{\rm A}$, and for the ones with $x^+/x^-$ being a phase  $x= x_{\rm B}$. Then the Hubbard and the $\su(2|2)$ spin chain model's parameters correspond to $\lam=+\infty$.}

Since $y+1/y=2v$ for any given $v$ one has two possible corresponding $y$-roots
and they can be parametrized as
\be\la{yr}
y_+=x(v)\,,\quad y_-={1\ov x(v)}\,, 
\ee
so that the set of  complex $y$-roots is divided into these two subsets.
Here and in what follows $x(v) = x_{\rm A}(v)$ for the  Hubbard  and A-models, and $x(v) = x_{\rm B}(v)$ for 
the B-model.
The two distinct types of $y$-roots 
$y_\pm$ 
obviously lead to two 
types $p_\pm(v)$ and $\E_\pm(v)$ of the momentum $p$ and the energy $\E$ as functions of $v$, see  \eqref{pk} and \eqref{disprels}  respectively. 

\medskip

The  string hypothesis, see \cite{Takahashi,Takbook,book,AF09a}, states that each root of the Bethe equations in the thermodynamic limit is a member of the following types of Bethe strings:
\begin{enumerate}
\item $y$-particle: a single charge  spin-up excitation parametrized by its real momentum $p$. 
The reality of $p$ then implies that 
\be\la{yp}
\begin{aligned}
&|y|=1\,,\quad |v|\le 1\,,\quad v\in {\mathbf R}\ \ {\rm for\ \ Hubbard\ and\ A-models}\,,\\\
&y\in {\mathbf R}\,,\quad |v|\ge 1\,,\quad v\in {\mathbf R}\ \ \ {\rm for\ \   B-model}\,.
\end{aligned}
\ee
Since $v$ is on the cut of the corresponding function $x$
its $y_\pm$ roots  can be also written as 
\bea\la{ypm}
y_\pm=x(v\pm i0)\,.
\eea
This formula  implies that for single charge excitations
$p_\pm(v)$ and  $\E_\pm(v)$ can be thought of as the values of the functions 
$p(v)\equiv p_+(v)$ and $\E(v)\equiv \E_+(v)$ on the upper and lower edges of the cut of $x(v)$. One finds for all the models that moving around the cut of $x(v)$ in the counter-clockwise direction increases the momentum $p$, and, therefore, $dp_-/dv>0$, $dp_+/dv<0$ for $v$ being
on the cut.

\item M$|{vw}$-string: 
a spinless charge $2M$ bound state composed of
 $2 M$ roots $y_j$ and $M$ roots $w_j$ and parametrised by $v\in {\mathbf R}$
\bea
v_j &=&v+(M+2-2j)\,\uh\, i,\quad v_{-j}=v-(M+2-2j)\,\uh\, i,\\
w_j &=&v+(M+1-2j)\,\uh\, i,\quad  j=1,\ldots,M\,,
\eea
where the roots $y_j$ are expressed through $v_j$ as 
\be\la{yMvw}
y_{j}=x(v_{j})\,,\quad  y_{-j}=1/x(v_{-j})\,,\quad j=1,\ldots ,M\,.
\ee
As a result, the momentum and the dispersion relation for the M$|{vw}$-string are 
\be\la{pEvw}
p_{M|vw}(v)=\sum_{j=1}^{M} \left[p_+\left(v_j\right)+p_-\left(v_{-j}\right)\right]\,,\quad \E_{M|vw}(v)=\sum_{j=1}^{M} \left[\E_+\left(v_j\right)+\E_-\left(v_{-j}\right)\right]\,.~~
\ee
One can check for all the models that $dp_{M|vw}/dv<0$ for any real $v$.

\item M$|{w}$-string: a spin $-M$ bound state composed of
$M$ roots $w_j$ and parametrised by $w\in {\mathbf R}$
\bea
w_j=w+(M+1-2j)\,\uh\, i,\quad  j=1,\ldots,M.
\eea
This family includes the $1|w$-string which has a single real root $w$.
\end{enumerate}
The string hypothesis for the Hubbard and A-models differs from the one for the B-model only by the location of a root $y$ of a single charge excitation. This  
is accommodated by the 
functions $x_{\rm A,B}(v)$, and in terms of  these
all the formulae look the same. 

\subsection{Free energy and TBA equations}

In the thermodynamic limit we introduce the densities $\rho_k$ and $\bar\rho_k$ of particles and holes,  and the 
Y-functions $Y_k   \equiv \bar\rho_k/\rho_k$ and dressed energies $\epsilon_k \equiv T\log Y_k$, where $k=\pm\,, M|vw\,, M|w\,,\ M\ge 1$. Since, as was discussed above, the roots $y_\pm$ 
are equal to the values of $x(v)$ on the upper and lower edges of its cut, 
the  
$Y_\pm$-functions satisfy the same property and can be viewed as 
the values of the function $Y(v)$ on the upper and lower edges of its square-root branch cut coinciding with the cut of $x(v)$
\bea\la{Ypm}
Y_+(v)=Y(v+i0)\,,\quad Y_-(v)=Y(v-i0)\,.
\eea

The derivation of the free energy and TBA equations follows a textbook route \cite{book}, see appendix \ref{dervTBA} for some detail. The free energy is given by
\bea\la{freng}
&&f=-\mu+\,\frac{1}{2\pi}(\E -\E_{1|vw}\star s){\circledast}\frac{{\rm d}\,p}{{\rm d}v} -\frac{T}{2\pi}\log\left(1+Y\right){\circledast}\left(\frac{{\rm d}\,p}{{\rm d}v}-s\star \frac{{\rm d}\,p_{1|vw}}{{\rm d}v} \right)~~~~~~~~\\\notag
&&
 \hspace{8.5cm} -\frac{T}{2\pi}\log\left(1+Y_{1|w}\right)\star s\,{\circledast}\frac{{\rm d}\,p}{{\rm d}v}\, ,
 \eea
where 
the symbol $\star$ denotes the following ``convolution'' 
\be\la{star}
g\star h\equiv \int_{-\infty}^\infty \, dt\, g(u,t)h(t,v)\,, 
\ee
and
 $\circledast$ denotes the contour integral in the counter-clockwise direction around the branch cut of $Y(v)$ or, equivalently, of $\E(v)$ and $p(v)$ located on the real axes. 
 Explicitly for the Hubbard and A-models one has
\bea\la{star2}
g\circledast  h =\int_{|t|\le 1} \, dt\,\left( g_-(u,t)h_-(t,v) - g_+(u,t)h_+(t,v)\right) = 
g_-\hstar h_--g_+\hstar h_+\,,~~~~~
\eea
while for the B-model
\bea\la{star3}
g\circledast  h =\int_{|t|\ge 1} \, dt\,\left( g_-(u,t)h_-(t,v) - g_+(u,t)h_+(t,v)\right)= 
g_-\cstar h_--g_+\cstar h_+\,,~~~~~
\eea
where 
\bea
g_\pm(u,t)\equiv g(u,t\pm i0)\,,\quad h_\pm(t,v)\equiv h(t\pm i0,v)\,,
\eea
and $\hstar$ and $\cstar$ denote convolutions with the integration over $|t|\le1$  and $|t|\ge1$ respectively.
If $g$ (or $h$) is a function of a single variable then one just drops $u$ (or $v$ or both) in (\ref{star}-\ref{star3}), e.g.  if $g=g(t)$ then $g\star h\equiv \int_{-\infty}^\infty \, dt\, g(t)h(t,v)$.
However, 
if  $g$ or $h$  is a kernel  defined through a function of one variable then it should be understood as $g(t,v)\equiv g(t-v)$.
Finally in \eqref{freng}  $s$ denotes a kernel defined in \eqref{skern}.

\medskip

The $Y$-functions  in \eqref{freng} are determined from the simplified TBA equations
\bea\la{stbay}
 \log Y(v)& =& \frac{\E -\E_{1|vw}\star s}{T}+\log\frac{1+Y_{1|vw}}{1+Y_{1|w}}\star s\,,\\ \la{stbayvw}
\log Y_{M|vw}(v) &=& I_{MN}\log\left(1+Y_{N|vw}\right)\star s -\delta_{M1} \log(1+Y){\circledast} s\,,\\\la{stbayw}
\log Y_{M|w}(v) &=& I_{MN}\log\left(1+Y_{N|w}\right)\star s -\delta_{M1} \log (1+\frac{1}{Y}){\circledast} s\,,
\eea
which are the conditions that the free energy is indeed minimised. 
Here $ I_{MN}=\delta_{M+1,N}+ \delta_{M-1,N}\,,\ M\ge2 $ and $ I_{1N}=\delta_{2N}$. Note that eq.\eqref{stbay} implies the following simple relation between $Y_\pm$ functions
\bea\la{yyEE}
 \log \frac{Y_+}{Y_-}&=&\frac{\E_+-\E_-}{T}\,,
\eea
following from the analyticity of the kernel $s(t-v)$ for real $v$.

In addition the TBA equations are supplemented with the asymptotics for large $M$
\bea
\lim_{M\rightarrow\infty} \frac{\log Y_{M|vw}}{M}=-\frac{2\,\mu}{T}\,,\qquad 
\lim_{M\rightarrow\infty} \frac{\log Y_{M|w}}{M}=\frac{2\,B}{T}\,,\la{Masym}
\eea
which follow from the canonical TBA equations, see appendix 
\ref{dervTBA} for the details. Note that the chemical potential $\mu$ and magnetic field $B$ enter the simplified TBA equations only via the large $M$ asymptotics of $Y_{M|vw}$ and $Y_{M|w}$.

It often happens that the sign of the final terms in equations \eqref{stbayvw}, \eqref{stbayw} play an important role in understanding the physics of the models. Through equation \eqref{yyEE} these signs follow from the relative magnitude of $\E_+$ and $\E_-$. Let us thus note that
\be\la{Epm}
\begin{aligned}
\E_--\E_+ &= -4\sqrt{1-v^2} \leq 0\quad\mbox{~~~~\,for the Hubbard model,}\\
\E_--\E_+ &= \frac{4 \uh^2
   \sqrt{1-v^2}}{\uh^2+v^2}\geq 0\quad\mbox{ ~~\,for the A model,}
   \\
\E_--\E_+ &= \frac{4 v^2 \sqrt{1-\frac{1}{v^2}}
  }{\uh^2+v^2}\geq 0\quad\mbox{ ~~for the  B model.}
\end{aligned}
\ee

In the case of the Hubbard model one can easily check that the TBA equations  and the free energy match exactly the ones in \cite{book}. To be precise one finds the  relations
\bea
&&e_0=\uh+\frac{1}{2\pi}(\E -\E_{1|vw}\star s){\circledast}\frac{{\rm d}\,p}{{\rm d}v} \,,\quad 
\s_0(v)=\frac{1}{2\pi}s\,{\circledast}\frac{{\rm d}\,p}{{\rm d}v}\,,\quad \\
&&
\rho_0\left(p(v)\right)\frac{{\rm d}\,p}{{\rm d}v}=\frac{1}{2\pi}\Big(\frac{{\rm d}\,p}{{\rm d}v}-s\star \frac{{\rm d}\,p_{1|vw}}{{\rm d}v} \Big)  \,,
\eea
where $e_0$, $\s_0$ and $\rho_0$ are the ground state energy per site,
the density of $w$-strings, and the density of roots of  single charge excitations for the half filled repulsive Hubbard model, respectively, see eqs.(5.69) and (5.70) of \cite{book}.

Finally it is worth pointing out that, as follows from the TBA equations, the Y-functions have  infinitely-many square-root branch points on the $v$-plane. To be precise,  $Y(v)$ has branch points  located at $v =\pm 1 + 2k\,\uh\, i\,, \ k \in {\mathbf Z}$, while $Y_{M|vw}$ and $Y_{M|w}$ have branch points  located at $v =\pm 1 + M\,\uh\, i + 2k\,\uh\, i$ and $v =\pm 1 - M\,\uh\, i - 2k\,\uh\, i\,, \ k \in {\mathbf N}$.

\section{Various limits}\la{varlims}

The free energy \eqref{freng} together with the TBA equations \eqref{stbay}-\eqref{stbayw} and their large $M$ asymptotics \eqref{Masym} contain all the information about the equilibrium states of the Hubbard, A- and B-models in the thermodynamic limit. In general one needs to numerically solve the coupled integral TBA equations to extract this information. In various limits however the equations simplify and one can make analytic progress. 

We first consider the weak and strong coupling limits.
As the kernel $s(v)$ 
has explicit dependence on $\uh$, it is useful to rescale the variable $v$ to $v/\uh$ and introduce a notation for functions that take a rescaled argument
\bea\la{rescale}
{\cal Y}_k(v)=Y_k(\uh\,v)\,,\quad{\mathfrak s}(v)={1\ov 4\,\cosh {\pi v\ov  2} }\,,\quad {\cal K}_M(v)=\frac{1}{\pi}\frac{M}{M^2 + v^2}\,.
\eea
Here we have also altered $s(v)$ and $K_M(v)$ by a factor of $\uh$ for convenience as they generally appear together with ${\rm d}v$. 
 Note that after rescaling the variable in the TBA equations they take the same form with Y-functions replaced $Y_k$ by  ${\cal Y}_k$ and $s$ by ${\mathfrak s}$. 
 
The limits of strong magnetic field, large chemical potential and infinite temperature are also discussed. Let us here introduce the convenient notation
\be
\beta = \frac{1}{T}\,,\quad \bc = \frac{\mu}{T}\,,\quad \bs = \frac{B}{T}\,.
\ee

\subsection{Weak coupling limit $\uh\to 0$}\la{weaklimit}

\subsubsection*{Hubbard and B model: free fermions}
From 
the Hamiltonians of the Hubbard model \eqref{HHub} and the B model \eqref{hamB} it is clear that they reduce to
free fermions in the weak coupling $\uh\rightarrow 0 $ limit. Here the limit is analysed on the level of the TBA equations. We find  as expected  that TBA equations can be solved explicitly and thus a closed expression for the free energy can be obtained.

 Let us consider the $\uh\rightarrow 0$ limit without rescaling the variable $v$.\footnote{It is worth pointing out that if in the Hubbard case one would rescale $v$ then the naive $\uh\rightarrow 0$ limit would lead to a constant solution of the TBA equations and to divergent integrals in the  free energy expression \eqref{freng}, while in the B case one would loose all the information about $Y_\pm$-functions.} First note that 
$
\lim_{\uh\rightarrow 0} s(v) = \frac{1}{2}\delta(v)
$. Next for $v$ on the intervals $\I^{\rm A}=(-1,1)$ and $\I^{\rm B}=(-\infty,-1)\cup(1,\infty)$ for the Hubbard and B-models respectively, one finds the following limits
\bea\la{epmdp}
&& \lim_{\uh\rightarrow0}\E_+(v)=\Delta\,,\quad \lim_{\uh\rightarrow0}\E_-(v)=-\Delta\,,\quad
 \lim_{\uh\rightarrow0}\E_{1|vw}(v)=2 \Delta\,,\\\nonumber
&& \lim_{\uh\rightarrow0} \frac{{\rm d} p_+}{{\rm d} v}(v)=-\frac{2}{\Delta}\,,\quad \lim_{\uh\rightarrow0}\frac{{\rm d} p_-}{{\rm d} v}(v)=\frac{2}{\Delta}\,,\quad
 \lim_{\uh\rightarrow0} \frac{{\rm d} p_{1|vw}}{{\rm d} v}(v)=-\frac{4}{ \Delta}\,,
\eea
where for the Hubbard model $\Delta=2\sqrt{1-v^2}$ and for the B-model $\Delta=-2\sqrt{1-\frac{1}{v^2}}$. Then 
for these values of $v$ 
the TBA equations become  a set of algebraic equations
{\small\bea\la{u0TBAp}
 \log Y_+ & =& \frac{1}{2}\log\frac{1+Y_{1|vw}}{1+Y_{1|w}}\,,\quad
\log {Y_+\ov Y_-}  = \frac{2\Delta}{T} \,,\\\la{u0TBAvw}
\log Y_{M|vw} &=& \frac{1}{2}I_{MN}\log\left(1+Y_{N|vw}\right) +\frac{1}{2}\delta_{M1} \log\frac{1+Y_+}{1+Y_-},\\\la{u0TBAw}
\log Y_{M|w} &=& \frac{1}{2}I_{MN}\log\left(1+Y_{N|w}\right) +\frac{1}{2}\delta_{M1} \log \frac{1+\frac{1}{Y_+}}{1+\frac{1}{Y_-}} \,.
\eea}Note that to compute the free energy \eqref{freng} one needs to know values of Y-functions  on $\I^{\rm A}$ and $\I^{\rm B}$ only.
The general 
solution to these algebraic TBA equations is well-known
and given by \eqref{Yinfty}.

In the limit $\uh\rightarrow0$ the free energy \eqref{freng} simplifies dramatically
\bean
f&=&-\mu-\frac{2}{\pi}\int_\I {\rm d}v    -\frac{T}{\pi}\int_{\I}\frac{{\rm d}v}{\Delta} \,\log\left((1+Y_-)\sqrt{1+Y_{1|w}}\,\right)\,,
\eean
and by using \eqref{Yinfty} one finds
\begin{align}
 \nonumber
 f=-T\int_{-\pi}^\pi \frac{{\rm d}p}{2\pi} \,\log\left((1+e^{\frac{2 \cos p +\mu+B}{T}})(1+e^{\frac{2 \cos p +\mu-B}{T}})\right)\,.
\end{align}
This is the expected result for free electrons \cite{Takbook}.

\subsubsection*{A-model: $\su(2|2)$ spin chain}

The weak coupling limit of the A-model is studied in detail in section \ref{TBAext} in the more general context of extended models. 
Here we point out that  the limit can be taken at the level of the TBA equations.  
However if one tries to take the limit keeping $v$ fixed then one finds that most of the  functions 
in \eqref{epmdp} exhibit singular behaviour at $v=0$. This can be resolved by rescaling $v$ to $v/\uh$.
 Note that then the  cut  $(-1,1)\rightarrow(-\infty,\infty)$ as $\uh\rightarrow0$ and thus ${\cal Y}_{\pm}(v)$ can be considered as distinct functions defined on ${\mathbb R}$. In this way the TBA equations \eqref{TBAexmod} with $\gh=0$ can  be obtained without going back to the Bethe ansatz for the $\su(2|2)$ spin chain.
 
\subsection{Strong coupling limit $\uh\to\infty$}\la{strongcoupling}

\subsubsection*{Hubbard and A-model: $\su(2)$ spin chain}
Here we consider the models in the limit of infinite coupling. The analysis is quite similar for the Hubbard and A-models and so these are examined together. 

 It is well known that the 
 strong coupling limit 
 of 
 the less than half-filled Hubbard model is the  {\it t-J}  model
(see appendix \ref{tJB} for a discussion), 
which at half-filling takes the form of 
the antiferromagnetic XXX spin chain
\bea\la{XXXsc}
{\mathbf H} = -2\uh -J\sum_{j=1}^{L} \left(\vec{\mathbf{S}}_j . \vec{\mathbf{S}}_{j+1} - \frac{1}{4} \right) - 2B\,{\mathbf S}^z\,,
\eea
where for later convenience we use $J=-\frac{1}{\uh}$.
The free energy and TBA equations for this Hamiltonian are known to be \cite{Takbook}:
\be\la{XXXTBA}
\begin{aligned}
f &= -2\uh+J\log 2 - T\int_{-\infty}^{\infty}{\rm d}v\,\log (1+\Y_1(v))\,{\mathfrak s}(v)\,,\\
\log \Y_M(v) =& I_{MN}\log\left(1+\Y_{N}\right)\star {\mathfrak s}\,\big(v\big) +\delta_{M1} \frac{2 \pi J  \,{\mathfrak s}(v)}{ T} \,,\quad
\lim_{M\rightarrow\infty} \frac{\log \Y_M}{M}=\frac{2 B}{T}\,.
\end{aligned}
\ee
The Hubbard free energy and TBA equations reduce to these as $\uh\rightarrow \infty$. 

Likewise we will see that in the strong coupling limit the free energy and TBA equations of the A-model reduce to \eqref{XXXTBA} with $J=\frac{1}{\uh}$. Thus it is natural to conclude that the strong coupling limit of the half-filled A-model is the ferromagnetic XXX spin chain with Hamiltonian given by \eqref{XXXsc} with   $J=\frac{1}{\uh}$.

First as $\uh\rightarrow \infty$ note that $\E^H\rightarrow \E^A\rightarrow -2\uh$,  $\E^H_{1|vw}\star s\rightarrow \frac{\log 2}{\uh}$ and $\E^A_{1|vw}\star s\rightarrow -\frac{\log 2}{\uh}$. Then from TBA equation \eqref{stbayvw} it follows that  $Y\rightarrow 0$. Furthermore 
$$\log(1+\frac{1}{Y}) {\circledast} s =\log{1+\frac{1}{Y_-}\ov  1+\frac{1}{Y_+}} {\hstar} s\ \rightarrow\ \frac{1}{T} (\E_+-\E_-)\hstar s$$ 
by equation \eqref{yyEE}.  The free energy and TBA equations for the Hubbard and A-models thus simplify to
\bea\la{TBAi}
&f =-\mu-2\uh+\kappa\frac{\log 2}{\uh} -\frac{T}{2\pi}\log\left(1+Y_{1|w}\right)\star s\,{\circledast}\frac{{\rm d}\,p}{{\rm d}v}\,,\\\la{TBAii}
&\log Y_{M|w}(v)  =  I_{MN}\log\left(1+Y_{N|w}\right)\star s -  \frac{\delta_{M1}}{T}(\E_+-\E_-)\hstar s \,,
\eea
with $\kappa = -1$ for the Hubbard model and $\kappa=1$ for the A-model. Since $\mu$ enters only the first term of the free energy it follows that the electron density is one, i.e. it is at half-filling.  These resemble quite closely the corresponding equations for the XXX spin chain \eqref{XXXTBA}.
Before  the exact equivalence is shown let us remark that the sign of the final term in \eqref{TBAii} gives an indication of whether one should expect an antiferromagnetic or ferromagnetic spin chain. From \eqref{Epm} we thus expect the Hubbard model to exhibit antiferromagnetic behaviour and the A-model to exhibit ferromagnetic behaviour. 

Next note the strong coupling limits 
\be\notag
\int_{-1}^1 {\rm d}t\, (\E_+(t)-\E_-(t))s(t,v) \rightarrow 
\Big\{ 
\begin{array}{ll}
 2\pi\,s(v)  & \mbox{for the Hubbard model,}   \\
 -2\pi\,s(v)  & \mbox{for the A-model.}        
\end{array} 
\ee
 Rescaling the kernel $s$  and Y-functions as in \eqref{rescale}, eq.\eqref{TBAii} takes the form
\bean 
{\cal Y}_{M|w}(v)  =  I_{MN}\log\left(1+{\cal Y}_{N|w}\right)\star {\mathfrak s}\,\big( v\big) +\kappa\, \delta_{M1} \frac{2 \pi  \,{\mathfrak s}(v)}{\uh\, T}\,.
\eean
To take into account the rescaling in the final term of equation \eqref{TBAi} note
\bean
&\log\left(1+Y_{1|w}\right)\star s\,{\circledast}\frac{{\rm d}\,p}{{\rm d}v} = \int_{-\infty}^{\infty}{\rm d}t\, \log\left(1+{\Y}_{1|w}(t)\right)\int_{-1}^{1}{\rm d}v\,  {\mathfrak s}(t,\uh v)\left(\frac{{\rm d}\,p_-}{{\rm d}v}(v)-\frac{{\rm d}\,p_+}{{\rm d}v}(v)\right)\,, \\
&\int_{-1}^{1}{\rm d}v\,  {\mathfrak s}(t,\uh v)\left(\frac{{\rm d}\,p_-}{{\rm d}v}(v)-\frac{{\rm d}\,p_+}{{\rm d}v}(v)\right)\xrightarrow{{\rm large }\,\uh}   2\pi {\mathfrak s}(t)\,.
\eean
Hence under the identification $\Y_{M|w}\equiv{\cal Y}_{M}$  the free energy and TBA equations for the Hubbard and A models do indeed reduce to those for the XXX spin chain \eqref{XXXTBA} with $J=-\frac{1}{\uh}$, $J=\frac{1}{\uh}$ respectively.

Let us comment on the attractive
versions of the models.  Here ${\tilde\E}^{\rm H}\rightarrow  \tilde\E^{\rm A}\rightarrow 2\uh$,  $\tilde\E^{\rm  H}_{1|vw}\star s\rightarrow - \frac{\log 2}{\uh}$, $\tilde\E^{\rm A}_{1|vw}\star s\rightarrow \frac{\log 2}{\uh}$ and so the TBA equation \eqref{stbayvw} imply $Y\rightarrow\infty$. It follows that the free energy (using the form \eqref{canfreng3}) and TBA equations simplify to
\bean
&f =-\mu+\kappa\frac{\log 2}{\uh} -\frac{T}{2\pi}\log\left(1+Y_{1|vw}\right)\star s\,{\circledast}\frac{{\rm d}\,p}{{\rm d}v}\,,\\
&\log Y_{M|vw}(v)  =  I_{MN}\log\left(1+Y_{N|vw}\right)\star s+  \frac{\delta_{M1}}{T}(\E_+-\E_-)\hstar s \,,
\eean
with $\kappa=-1$ for the attractive Hubbard model and $\kappa=1$ for the attractive A-model. Since there is no $B$ dependence in the equations the magnetisation vanishes. 
As the inequalities in \eqref{Epm} are reversed in going to the opposite models one sees that in the strong coupling limit the attractive Hubbard model is an antiferromagnetic XXX chain while the  attractive  A-model is a ferromagnetic XXX chain.
Note that these are charge $\su(2)$ chains, governed by ${\boldsymbol \eta}$, as opposed to spin chains. For the strongly attractive A-model one thus expects spontaneous symmetry breaking at $T=0$ with the model being empty for $\mu<0$ and having all sites doubly occupied for $\mu>0$.

\subsubsection*{B-model}

The strong coupling limit of the B-model is  studied in detail in section \ref{TBAext}. The analysis is similar to that for  the weak coupling limit of the A-model above. Note that if $v$ is rescaled to $v/\uh$ then the  cut  $(-\infty,-1)\cup (1,\infty)\rightarrow(-\infty,\infty)$ as $\uh\rightarrow\infty$. Again ${\cal Y}_{\pm}(v)$ can be considered as distinct functions defined on ${\mathbb R}$, and the TBA equations \eqref{TBAexmod} for the extended B-model with $\gh=0$ 
can be found. 

\subsection{Strong magnetic field limit $B\to\infty$}

In the limit $B\rightarrow \infty$ the Y-functions for $M|w$ strings diverge. This can be seen from the canonical TBA equations \eqref{TBAy}-\eqref{TBAw}. Note that the final term on the right hand side of equation \eqref{TBAw} is finite as $B\rightarrow\infty$ because $Y\rightarrow 0$ from \eqref{TBAy} and thus
\be\notag
\log (1+\frac{1}{Y}){\circledast} K_M\rightarrow -\E {\circledast} K_M\,,
\ee
using \eqref{yyEE}.
Therefore 
\be\notag
Y_{M|w} \to e^{2 M \bs},
\ee
and the free energy simplifies to
\bean
f=-\mu-B+\,\frac{1}{2\pi}(\E -\E_{1|vw}\star s){\circledast}\frac{{\rm d}\,p}{{\rm d}v}\,.
\eean
The electron density is one and the  magnetisation per site is one-half, and thus the ground state is full of spin-up electrons as expected.

\subsection{Large chemical potential limit $\mu\to-\infty$}

The analysis of the limit $\mu\rightarrow-\infty$ is 
similar to the limit $B\rightarrow \infty$. Here the  Y-functions for $M|vw$-strings diverge as follows from the canonical TBA equation  \eqref{TBAvw} and become 
\bean
Y_{M|vw} \to e^{-2 M \bc},
\eean
and from  \eqref{canfreng3} one finds the free energy 
\bean
f=\,\frac{1}{2\pi}\, \E_{1|vw}\star s {\, \circledast\,}\frac{{\rm d}\,p}{{\rm d}v}\,.
\eean
Thus the ground state is empty of all charge excitations. This is as expected as a large chemical potential means there is a large energy cost for having a particle in the system.

\subsection{Infinite temperature limit}
In the infinite temperature limit one expects many of the details of a system to become irrelevant. In particular we find that the Hubbard, A- and B-models become identical. 
In the limit
$T\rightarrow \infty$, with $\bc$ and $\bs$ fixed,
the driving term of the TBA equations \eqref{stbay}-\eqref{stbayw} becomes negligible. The Y-functions are thus constants and $Y_+=Y_-\equiv Y$. The TBA equations become recursion relations with the solution \cite{book}
\bean
Y=\frac{\cosh \bc}{\cosh \bs} \, , \quad Y_{M|vw}=\left[\frac{\sinh (M+1)\bc}{\sinh \bc}\right]^2-1\,,\quad Y_{M|w}=\left[\frac{\sinh (M+1)\bs}{\sinh \bs}\right]^2-1\,,
\eean
and the free energy   becomes
\bean
f= - \mu -T\log(1+Y)-\frac{T}{2}\log(1+Y_{1|vw})=-T\log\left(2 e^{\bc}\cosh \bs+e^{2 \bc}+1\right)\,.
\eean
In the strict infinite temperature limit $\bc\rightarrow 0$ and $\bs\rightarrow 0$. Thus $f\rightarrow -T\log 4$ as expected for a model with  four degrees of freedom per site.

\section{Zero temperature limit}\la{zeroT}

The $T\rightarrow 0 $ limit is of great interest as the absence of thermal fluctuations allows one to obtain a clear picture of the phase diagram. The TBA equations simplify dramatically and it is possible to study them analytically in certain regimes. Indeed in this limit it is useful to work with the canonical TBA equations \eqref{TBAy}-\eqref{TBAw}\footnote{In this section we use the subscript $y$ to label quantities related to $y$-particles.}
\bean
\log Y_y &=& \frac{\E_y-\mu-B}{T}+\log \frac{1+\frac{1}{Y_{N|vw}}}{1+\frac{1}{Y_{N|w}}}\star K_{N},~~~~~\la{Tbay2}\\
\log Y_{M|vw} &=& \frac{\E_{M|vw}-2\,M\,\mu}{T}-\log\left(1+\frac{1}{Y_y} \right)\circledast K_M+\log\left(1+\frac{1}{Y_{N|vw}}\right)\star K_{NM},~~~~~\la{Tbavw2}\\
\log Y_{M|w}  &=& \frac{2\,M\,B}{T}- \log\left(1+\frac{1}{Y_y} \right) \circledast K_M+\log\left(1+\frac{1}{Y_{N|w}}\right)\star K_{NM}.~~~~~\la{Tbaw2}
\eean
These equations are the direct result of the minimisation of the free energy and it is from them that the simplified TBA equations are derived, see appendix \ref{dervTBA}. They contain $\mu$ and $B$ explicitly rather than through the asymptotics \eqref{Masym} which will not be relevant for the discussion of zero temperature.  
Furthermore, the Y-functions will vanish and diverge as $T\rightarrow 0$ and so we will work with the dressed energies which remain finite
\bea
\e_k=T\log Y_k,
\eea
where $k$  labels the string type. Note then that
 \be\la{YT0s}\begin{array}{ccccl}
Y_k<1 &\Rightarrow& \e_k<0 &\Rightarrow & \lim_{T\rightarrow 0} Y_k=0\,,\\
Y_k>1 &\Rightarrow& \e_k>0 &\Rightarrow & \lim_{T\rightarrow 0} Y_k=\infty\,.
\end{array}
\ee
Recalling  $Y_k   \equiv \bar\rho_k/\rho_k$, the ratio of the density of holes to the density of particles, 
it follows that $\e_k(v)<0$ means that there are no holes of type $k$ with spectral parameter $v$ in the ground-state whereas $\e_k(v)>0$ means that there are no particles of type $k$ with spectral parameter $v$ in the ground-state. From the canonical TBA equations it is seen that a Y-function does not contribute if $Y_k(v)>1$ for all $v$, i.e. that there are no particles of that type.

It is possible to rule out a lot of string types by momentarily considering the simplified TBA equations  \eqref{stbay}-\eqref{stbayw}. Note that
\bean
Y_{M|vw}>1 \quad\mbox{and}\quad Y_{M|w}>1 \quad\mbox{for}\quad M\geq 2\,,
\eean
as the right hand sides of the equations for these strings are strictly positive. The sign of the extra term in the TBA equations for $Y_{1|vw}$ and $Y_{1|w}$ follows from the inequalities \eqref{Epm}. 
Thus $Y_{1|vw}>1$ for the Hubbard model as is well known, whereas for the A- and B-models one has that $Y_{1|w}>1$. 

With this insight the canonical TBA equations simplify dramatically at zero temperature. From now on we consider only the cases of the A- and B-models. Writing the equations in terms of the dressed energies they take the form
\be\la{T0TBA}
\begin{aligned}
\e_y&=\E_y-\mu-B-\e_{1|vw}\star_{Q_{1|vw}}\, K_1\,,\\
\e_{1|vw}&=\E_{1|vw}-2 \mu + \e_y \circledast_{Q_y}\, K_1 - \e_{1|vw}\star_{Q_{1|vw}}\, K_{2}\,,
\end{aligned}
\ee
where $K_{11}=K_2$ was used.
Here a subscript notational convention has been adopted for the convolutions to signify that the integrals should only be taken over the region for which $\e_k<0$, as for $\e_k>0$ one has that $Y_k\rightarrow \infty$ as in \eqref{YT0s} and the contribution vanishes. Here $Q_k$ is the interval defined by 
\bean
\e_k(v)<0 \quad\mbox{for}\quad v\in Q_k.
\eean
We see from the definition of Y-functions that $Q_{1|vw}\subseteq {\mathbb R}$ and $Q_y =Q_-\cup Q_+$,  where  $Q_\pm \subseteq \I^{\rm{A,B}}_\pm$ (recall that 
$\I^{\rm{A,B}}_\pm$ are the upper and lower edges of the cuts of $x^{\rm{A,B}}(v)$).

\medskip

 The phase diagram is a description of the constituent particles of the ground state as a function of $\mu$ and $B$. Recall that the Bethe ansatz deals with states in the regime $\mu\leq 0$ and $B\geq 0$ and so it is for this quadrant that the TBA equations provide information. The phase diagram for the other three quadrants follows straightforwardly  from the symmetries \eqref{frengsym} of the free energy.

 The ground state can consist of only $y$-particles (spin-up electrons) and $1|vw$ strings (bound states of a spin-up electron and a spin-down electron). 
There are five phases that will be relevant to our discussion:
\begin{enumerate}[label=\Roman{*}.]
\item  Empty band: no $y$-particles, no $1|vw$-strings.
\item Partially filled and spin polarised band: some $y$-particles, no $1|vw$-strings.
\item Half filled and spin polarised band: saturated with $y$-particles, no $1|vw$-strings.
\item Partially filled and partially spin polarised band: some $y$-particles, some $1|vw$-strings.
\item Half filled and partially spin polarised band: some $y$-particles, saturated with $1|vw$-strings.
\end{enumerate}
Further phases such as partially filled and spin neutral are not identified as such will correspond to points and lines rather than regions of the phase diagram.

Boundaries between phases are determined as follows: the ground state begins to contain a string of type $k$ when $\mu$ and $B$ are tuned so that $\min_v \e_k(v)=0$, and the ground state becomes saturated with a string of type $k$ when $\mu$ and $B$ are tuned so that $\max_v \e_k(v)=0$.

 \subsubsection*{A-model}
 
 In the A-model there are no $1|vw$-strings in the zero temperature ground state. This can be seen from a numerical study of the TBA equations \eqref{T0TBA}.  One finds that $\e_{1|vw}<0$ is not consistent with $B>0$ and so there can be no $1|vw$-strings. Thus the model can only exhibit phases I, II and III, and the relevant TBA equation is
 \bea\la{ATBA}
 \e_y=\E_y-\mu-B.
 \eea
 Recalling the dispersion relation from \eqref{disprels}, we see that the ground state is empty for $\mu<-2-2\sqrt{1+\uh^2}-B$, then partially filled up to the line $\mu=2-2\sqrt{1+\uh^2}-B$, and saturated with $y$-particles thereafter. The phase diagram for the $A$-model for $\mu\leq0$, $B\geq0$ is presented in Figure \ref{phasediags}.
Note that  $B=0$ is a singular point at $T=0$ arising from spontaneous breaking of $\su(2)_s$.

\subsubsection*{B-model}
 In the B-model there may exist both $y$-particles and $1|vw$-strings in the ground state.  Let us determine the phase diagram on a case by case basis.
 
First consider the question of when $1|vw$-strings begin to appear in the ground state, this determines the boundary between phase IV and phases I, II and III. When there are no $1|vw$-strings then $\e_{1|vw}(v)\geq0$ for all $v$. Thus the convolutions $\e_{1|vw}$ do not contribute to the right hand side of the TBA equations \eqref{T0TBA}, and they become
\be\la{BT0TBA}
\begin{aligned}
\e_y&=\E_y-\mu-B,\quad
\e_{1|vw}=\E_{1|vw}-2 \mu + \e_y \circledast_{Q_y}\, K_1.
\end{aligned}
\ee
The phase boundary is determined by the condition that $\min_v \e_{1|vw}(v)=0$.
Note that $\e_y(v)$ is even and a monotonically increasing function of $v$ around the cut $(1,\infty)$ from the minimum of $\E_y(v)$ at $v=\infty +i0$ to the maximum of $\E_y(v)$ at $v=\infty -i0$, that is  $\e_+(v)$ and $\e_-(v)$ are decreasing and increasing  for positive $v$ respectively. 
Also  $\e_{1|vw}(v)$ is even and a monotonically decreasing function\footnote{This second property is seen via the identity
$
 \E_{1|vw}=-\E_y\circledast K_1= -\e_y\circledast K_1 ,
$
 where the first equality follows from shifting the contours of integration and the second from the fact that the constant terms in $\e_y$ cancel when integrated over both edges of the cut. This implies that
$
 \e_{1|vw}=-2\mu-\e_y \circledast_{{\bar Q}_y}\, K_1,
$
 where ${\bar Q}_y$ is the complement of $Q_y$ in the contour  $\I^{\rm B}_+\cup \I^{\rm B}_-$ around the cut, and it follows that the derivative of $\e_{1|vw}(v)$ is negative for $v\in(0,\infty)$.} 
 for positive $v$. 
 Thus the minimum of $\e_{1|vw}(v)$ is at $v=\pm\infty$. By evaluating the dressed energies at $\pm\infty$ one can see that they have finite asymptotics, denoted by $\e_k^\infty$. To do so requires the knowledge that for a function $f(v)$ which asymptotes to $f^\infty$ one has
 \bean
 \lim_{v\rightarrow\infty} \int_{\Lambda}^{\infty}{\rm d}t\,f(t)\,K_1(t,v)=\lim_{v\rightarrow\infty} \int_{\Lambda-v}^{\infty}{\rm d}t\,f(t+v)\,K_1(t,0)=f^\infty,
 \eean
 for any finite $\Lambda$. These can be used to determine the phase boundaries for the following cases  i)  $Q_y=\o$, the boundary between phases I and IV, ii) $Q_y \subset  \I^{\rm B}_+\cup \I^{\rm B}_-$, a proper subset, the boundary between phases II and IV, iii) $Q_y= \I^{\rm B}_+\cup \I^{\rm B}_-$, the boundary between phases III and IV:
\begin{enumerate}[label=\roman{*})]
\item  Evaluated at $v=\infty$ the TBA equations \eqref{BT0TBA} reduce to the linear equations
\bean
\e_+^\infty = -2-\mu-B,\quad
\e_-^\infty = 2-\mu-B,\quad
\e_{1|vw}^\infty = -4-2 \mu.
\eean
From the conditions $\e_+^\infty\geq0$ and $\e^\infty_{1|vw}=0$ it follows that the boundary of phases I and IV is the point $(\mu,B)=(-2,0)$.
\item  Evaluated at $v=\infty$ the TBA equations \eqref{BT0TBA} reduce to the linear equations
\bean
\e_+^\infty = -2-\mu-B,\quad
\e_-^\infty = 2-\mu-B,\quad
\e_{1|vw}^\infty = -4-2 \mu-\e_+^\infty.
\eean
From the condition $\e^\infty_{1|vw}=0$ it follows that the boundary of phases II and IV is the line $B=2+\mu$.
\item  Evaluated at $v=\infty$ the TBA equations \eqref{BT0TBA} reduce to the linear equations
\bean
\e_+^\infty = -2-\mu-B,\quad
\e_-^\infty = 2-\mu-B,\quad
\e_{1|vw}^\infty = -4-2 \mu+4.
\eean
From the conditions $\e^\infty_-\leq0$ and $\e^\infty_{1|vw}=0$ it follows that the boundary of phases III and IV is the point $(\mu,B)=(0,2)$.
\end{enumerate}

Next let us determine the phase boundaries for $y$-particles. Neglecting for a moment the contribution from $1|vw$-strings, the relevant TBA equation takes the form
 \bean
 \e_y=\E_y-\mu-B.
 \eean
 Then recalling the dispersion relation \eqref{disprels}, in a similar fashion to the A-model, one can see that the line $\mu=-2-B$ would separate phases I and II and the line $\mu=2-B$ would separate phases II and III. Note that neither of these lines enter the region,  located in the previous paragraph, for which $1|vw$-strings appear in the ground state. Thus neglecting the $1|vw$-strings in the determination of the $y$-particle phase boundaries turns out to be self-consistent. 

Finally consider when the ground state becomes half-filled. Let us begin by  establishing that this happens when $Q_{1|vw}={\mathbb R}$. The number of $y$-particles per site is $n_y=1\,\cstar_{Q_+} \,\rho_+\, +1\,\cstar_{Q_-}\, \rho_-$, the number of $1|vw$-strings per site is $n_{1|vw}=1\,\star_{Q_{1|vw}}\, \rho_{1|vw}$, and then the total number of electrons per site is $n_c=n_y+2n_{1|vw}$.  Consider the zero temperature equation for the density of $1|vw$-strings \footnote{The zero temperature limit of the equations for densities is discussed in appendix \ref{dervTBA}, see (\ref{tbaygr0}).}
\bean
\rho_{1|vw}=-\frac{1}{2\pi}\frac{{\rm d}p_{1|vw}}{{\rm d}v}-K_1\cstar_{Q_+} \rho_+\,-K_1\cstar_{Q_-} \,\rho_-\, -K_2\star_{Q_{1|vw}}\rho_{1|vw}.
\eean
Note that $Q_{1|vw}={\mathbb R}$ implies
\bean
n_{1|vw}=1\star \rho_{1|vw} &=& 1 -1\star K_1\cstar_{Q_+} \rho_+\,-1\star K_1\cstar_{Q_-} \,\rho_-\,-1\star K_2\star \rho_{1|vw}\\
&=& 1- n_y - n_{1|vw},
\eean
and hence $n_c=1$ and the ground state is  half-filled. It is clear that $Q_{1|vw}\subset{\mathbb R}$, a proper subset, implies  $n_c<1$. Thus the boundary between phases IV and V is determined by setting $Q_{1|vw}={\mathbb R}$ and varying $Q_y$. 
First consider the case $Q_y=\o$. Here $\e_+\ge 0$  and evaluating the TBA equations \eqref{T0TBA} at $v=\infty$ one finds that $\e_+^\infty = -B$, and thus $B=0$.
Therefore the boundary is the point $(\mu,B)=(\mu_0,0)$ where $\mu_0$ is the solution to the equation $\e_{1|vw}(0)=0$, with $\e_{1|vw}$ determined  by the integral equation
\bean
 \e_{1|vw} = \E_{1|vw}-2\mu- \e_{1|vw}\star K_2\,.
\eean
Using the identities $s+K_2\star  s=K_1$ and $\E_{1|vw}=-\E_y  \circledast \, K_1$ this equation can be inverted 
 \bea\la{mu0}
 \mu_0 = -\big( \E_y\circledast s \big)(0)= -\int_{|t|>1}\, dt\, \frac{4 t^2 \sqrt{1-\frac{1}{t^2}}
  }{\uh^2+t^2}\,s(t) \,.
 \eea
  Evaluating this integral one finds that $\mu_0$ ranges from $\mu_0 = 0$ at weak coupling to $\mu_c=2\log 2 -2\approx -0.6137$ at strong coupling. Next for  the case $Q_y  =\I^{\rm B}_+\cup \I^{\rm B}_-$\,, examining the asymptotics of the TBA equations as above yields that the boundary is the point $(\mu,B)=(0,2)$. Thus phases II, III, IV and V all meet here. Furthermore it follows  that at this point $\e_{1|vw}(v)=0$ for all $v$ and hence there are neither particles nor holes of $1|vw$ strings in the ground state here. To fill in the picture the line joining $(\mu_0, 0)$ and $(0,2)$ must be determined numerically from the coupled integral equations
\bean
\e_y=\E_y-\mu-B-\e_{1|vw}\star\, K_1,\quad
\e_{1|vw}=\E_{1|vw}-2 \mu + \e_y \circledast_{Q_y}\, K_1 - \e_{1|vw}\star\, K_{2}\,,
\eean
parametrised by the interval $Q_y  \subset \I^{\rm B}_+\cup \I^{\rm B}_-$.
A practical way to achieve this is to introduce the derivative with respect to $v$ of these equations
\bean
\e'_y=\E'_y - \e'_{1|vw}\star\, K_1,\quad
\e'_{1|vw}=\E'_{1|vw} + \e'_y \circledast_{Q_y}\, K_1 - \e'_{1|vw}\star\, K_{2},
\eean
where here we have used $\frac{\partial }{\partial v}K_M(t-v) = -\frac{\partial }{\partial t}K_M(t-v) $, integrated by parts, and used that $\e_y(v)=0$ on the boundary of $Q_y$. Then for a given interval $Q_y$ the corresponding point $(\mu,B)$ is found by identifying the asymptotics $\e_+^\infty=-2B$, $\e_{1|vw}^\infty = -2 -\mu +B$ with
\bean
\e_+^\infty =\int_{Q^+, t>0} \,dt\,\e'_+(t) - \int_{Q^-, t>0} \,dt\,\e'_-(t)\,,\quad
\e_{1|vw}^\infty = \int_0^\infty {\rm d}t\, \e'_{1|vw}(t)\,.
\eean


\section{TBA for the extended models}\la{TBAext}
The TBA equations for the weak coupling limit of the A-model and the strong coupling limit of the B-model can be derived from the TBA equations (\ref{stbay}-\ref{stbayw}) as discussed in sections \ref{weaklimit}, \ref{strongcoupling}. Now we wish to investigate the integrable extensions of these models, presented in section \ref{exmods}. To this end we review the necessary modifications to the analysis of the previous four sections. The notation of writing in script kernels which do not depend on $\uh$ due to a rescaling of their argument is used, see \eqref{rescale}.

\subsection{Bethe and TBA equations}

Let us first consider the models without the extension, so $\gh=0$. 
To handle the Bethe equations \eqref{BE1} correctly 
 it is necessary to rescale $v\rightarrow v/\uh$, $w\rightarrow w/\uh$ and then take the appropriate limits of $\uh$. 
Then  they become\footnote{These Bethe equations coincide with those derived in \cite{EKS93} with the BFFB grading.}
\be\notag
\begin{aligned}
e^{i p_{+,k} L} =\Big(\a\, \frac{z_{+,k}+i}{z_{+,k}-i} \Big)^L =&
\prod_{j=1}^{M}\frac{z_{+,k}-w_j+i}{z_{+,k}-w_j-i}\,, & k=1,\ldots,N_+,\\
\prod_{j=1}^{N_+}\frac{w_k-z_{+,j}+i}{w_k-z_{+,j}-i} \prod_{j=1}^{N_-}\frac{w_k-z_{-,j}+i}{w_k-z_{-,j}-i}
=& \prod_{j=1,j\neq
k}^{M}\frac{w_k-w_i+2i}{w_k-w_j-2i}\,,& k=1,\ldots,M,\\
e^{i p_{-,k} L} =(-1)^L =& \prod_{j=1}^{M}\frac{z_{-,k}-w_j+i}{z_{-,k}-w_j-i},& k=1,\ldots,N_-,
\end{aligned}
\ee
where  $\a=-1$  for the extended A and opposite A models, and $\a=1$  for the extended B and opposite B models.
 The dispersion relations for the extended A- and B-models are 
 \bean
 \E^{\rm A_0}_+(v)=-2\cos p_+(v)-2=-4\pi{\cal K}_1(v)\,,\,\,  & \quad\,\,\, \, \E^{\rm A_0}_-(v)=-2\cos p_-(v)-2=0\,,   \\
  \E^{\rm B_\infty}_+(v)=-2\cos p_+(v)=4\pi{\cal K}_1(v)-2\,, \quad&\E^{\rm B_\infty}_-(v)=-2\cos p_-(v)=2 \,, 
\eean
 and the respective dispersion relations for the opposite models have the opposite sign.
In the limiting process the $y$-roots decouple into two sets
\bean
p_+:&& \Big\{y_+=x(\uh\, v): {\mathcal Im}(v)\geq 0\} \cup \{y_-={1\ov x(\uh\, v)}: {\mathcal Im}(v)< 0\}\,,\\
p_-:&& \{y_-={1\ov x(\uh\, v)}: {\mathcal Im}(v)\geq 0\}  \cup\{y_+=x(\uh\, v): {\mathcal Im}(v)< 0\}\,,
\eean
with the momenta as presented in the Bethe equations above. The corresponding $v$-roots are labelled $z_+$ and $z_-$. 
Note that the $z_-$-roots have been frozen,  while the $z_+$-roots retain non-trivial momentum dependence. As
the Hubbard term commutes with the respective Hamiltonians,
 the number of doubly occupied sites is conserved and should be  either $N_+$ or $N_-$. It is clear that it must be  $N_-$, the number of frozen roots. 
Indeed this is understood  as $z_+$-roots being a first level of electrons  that occupy empty sites, and that $z_-$-roots are an extra level that combine with a $z_+$-root and a $w$-root to  make a site doubly occupied. 
As a consistency check note that there are at most $M$ $z_-$-roots as each must satisfy the same polynomial of degree $M$, and as $2M\leq N_+ + N_-$, that for each $z_-$ root there exists a corresponding  $z_+$-root and $w$-root.

Next we make the observation that $M|vw$-strings are composed of $M+1$ $z_+$-roots and $M-1$ $z_-$-roots. Let us illustrate this for the case of the $1|vw$-string. Recalling equation \eqref{yMvw}, $y_1=x(v+i)$ and $y_{-1}={1\ov x(v-i)}$, it is clear that the two $y$-roots are of  $p_+$ type. For longer strings the logic extends naturally. 
This has an important implication. The $1|vw$-string here corresponds  to a bound pair of a spin-up electron and a spin-down electron, each on a singly occupied site. Indeed, the Hubbard interaction favours singly occupied sites and thus extending the models can favour electron pairing.

Now consider how the coupling constant $\gh$ enters the analysis. 
Since $z_+$ corresponds to a singly occupied site the Hubbard term shifts its dispersion relation as $\E_+\to \E_+ -2\gh$. Similarly
the dispersion relation for $z_-$ is shifted as $\E_-\to \E_- +2\gh$ because 
it corresponds to a doubly occupied site. Finally since an $M|vw$-string is composed of $M+1$ $z_+$-roots and $M-1$ $z_-$-roots its dispersion relation is shifted as $\E_{M|vw}\to \E_{M|vw} -4\gh$.

Despite apparent differences between the extended A- and B-models, 
 the Bethe equations above allow an exact equivalence to be seen.
 Note that they  are identical for lattices of even length $L$. Moreover shifting $\gh\rightarrow \gh-1$ takes the dispersion relations for the B-model to the dispersion relations for the opposite A-model. Similarly, shifting $\gh \rightarrow \gh + 1$ takes the dispersion relations for the opposite B-model to the dispersion relations for the A-model. Thus the spectra  of the related models are identical  for lattices of even length. At the level of the Hamiltonians \eqref{HamA0}-\eqref{opHamBinf}, this implies the following unitary equivalences  as discussed in section \ref{exmods}  
\be\la{EKSvsB}
{\mathbf H}^{\rm {\tilde A}_0}(\gh) 		  \sim  {\mathbf H}^{\rm B_\infty}(\gh-1) \,,\quad
 {\mathbf H}^{\rm A_0}(\gh)   \sim   {\mathbf H}^{\rm {\tilde B}_\infty}(\gh+1)\, .
 \ee
In particular it follows that the EKS model  and the extended B-model are the same up to a shift of the coupling $\gh$ by 1.

\smallskip

Now let us discuss the TBA equations.
The eigenvalue of the free energy, the analogue of equation \eqref{freval}, is
\be\la{freval2}
\begin{aligned}
f(\mu,B,T)&=\int {\rm d}v (\E_+ -\mu-B-2\gh )\rho_{+}+\int {\rm d}v (\E_-  -\mu-B+ 2\gh)\rho_{-}\\
&+\sum_{M=1}^{\infty}\int {\rm d}v (\E_{M|vw}  - 2 M\, \mu- 4\gh)\rho_{M|vw}+\sum_{M=1}^{\infty}\int {\rm d}v\, 2 M \,B \,\rho_{M|w}-T\,s,
\end{aligned}
\ee
with the dispersion relations for $M|vw$-strings given by
\bean
  \E^{\rm A_0}_{M|vw}(v)=-4\pi{\cal K}_{M+1}(v)\,, \quad   \E^{\rm B_\infty}_{M|vw}(v)=4\pi{\cal K}_{M+1}(v)-4\,,
\eean
and the signs are changed for the opposite models. The effect of $\gh$ is to shift the chemical potential, differently for different string types.
The canonical TBA equations are thus
\be
\begin{aligned}\la{TBAexmod}
\log \Y_{+} &= \frac{\E_+-\mu-B-2\gh}{T}+\log \frac{1+\frac{1}{\Y_{N|vw}}}{1+\frac{1}{\Y_{N|w}}}\star {\cal K}_{N},\\
\log \Y_{-} &= \frac{\E_- -\mu-B+2\gh }{T}+\log \frac{1+\frac{1}{\Y_{N|vw}}}{1+\frac{1}{\Y_{N|w}}}\star {\cal K}_{N},\\
\log \Y_{M|vw} &= \frac{\E_{M|vw} -2\,M\,\mu- 4\gh}{T}+\log\Big(1+\frac{1}{\Y_{N|vw}}\Big)\star {\cal K}_{NM}+ \log{1+\frac{1}{\Y_+} \ov 1+\frac{1}{\Y_-}}\star {\cal K}_M\,,\\
\log \Y_{M|w} &= \frac{2\,M\,B}{T}+\log\Big(1+\frac{1}{\Y_{N|w}}\Big)\star {\cal K}_{NM}+ \log{1+\frac{1}{\Y_+} \ov 1+\frac{1}{\Y_-}}\star {\cal K}_M\,.
\end{aligned}
\ee
Simplifying these equations as in appendix \ref{dervTBA} they become
\be\la{TBAexten}
\begin{aligned}
 \log \Y_+& = \frac{ \E_+ -\E_{1|vw}\star  {\mathfrak s}}{T}+\log\frac{1+\Y_{1|vw}}{1+\Y_{1|w}}\star  {\mathfrak s}\,,\\
 \log \Y_-& = \frac{ \E_- -\E_{1|vw}\star  {\mathfrak s}+4\gh}{T}+\log\frac{1+\Y_{1|vw}}{1+\Y_{1|w}}\star  {\mathfrak s}\,,\\
\log \Y_{M|vw} &= I_{MN}\log\big(1+\Y_{N|vw}\big)\star  {\mathfrak s} + \delta_{M1} \log{1+ \Y_+ \ov 1+ \Y_-} \star  {\mathfrak s}\,,\\
\log \Y_{M|w} &= I_{MN}\log\big(1+\Y_{N|w}\big)\star  {\mathfrak s} + \delta_{M1} \log{1+\frac{1}{\Y_+} \ov 1+\frac{1}{\Y_-}} \star {\mathfrak s}\,,~~~~~~~~~~~~~~~~~
\end{aligned}
\ee
with the familiar large $M$ asymptotics 
\bean
\lim_{M\rightarrow\infty} \frac{\log \Y_{M|vw}}{M}=-\frac{2\,\mu}{T}\,,\qquad 
\lim_{M\rightarrow\infty} \frac{\log \Y_{M|w}}{M}=\frac{2\,B}{T}\,.
\eean
The driving terms in these equations are simplified and  presented explicitly in Table \ref{tb2}.
\begin{table}[t]
\begin{center}
\begin{tabular}{c|c|c|c|c}
 & ${\rm A_0}$ & ${\rm \widetilde{A}_0}$& $\rm B_\infty$& $\rm \widetilde{B}_\infty$\\
\hline
$\E_+$ &$-4\pi{\cal K}_1$& $4\pi{\cal K}_1$ & $4\pi{\cal K}_1-2$ & $2-4\pi{\cal K}_1$
 \\
$\E_-$ &$0$& $0$ & $2$ & $-2$
 \\
 $\E_{1|vw}$ &$-4\pi{\cal K}_2$& $4\pi{\cal K}_2$ & $4\pi{\cal K}_2-4$ & $4-4\pi{\cal K}_2$
 \\
$\E_+ -\E_{1|vw}\star  {\mathfrak s}$ &  $-4\pi\,{\mathfrak s}$& $4\pi\,{\mathfrak s}$ & $4\pi\,{\mathfrak s}$ & $-4\pi\,{\mathfrak s}$ \\
$\E_- -\E_{1|vw}\star  {\mathfrak s}$ &$4\pi{\cal K}_1-4\pi\,{\mathfrak s}\,$& $4\pi\,{\mathfrak s}-4\pi{\cal K}_1$ & $4\pi\,{\mathfrak s}-4\pi{\cal K}_1+4$ & $4\pi{\cal K}_1\,-4\,\pi{\mathfrak s}-4$
 \\
$(J,c_A)$ &$(2,1)$& $(-2,-1)$ & $(-2,0)$ & $(2,0)$
\end{tabular}
\end{center}

\vspace{-0.5cm}

\caption{\small Driving terms of the TBA equations \eqref{TBAexten}, and constants for the free energy.}
\la{tb2}
\end{table}

Taking into account that
 \bean
 \frac{{\rm d} p_+}{{\rm d} v}= -2\pi{\cal K}_1(v)\,,\quad \frac{{\rm d} p_-}{{\rm d} v}=0\,,\quad
 \frac{{\rm d} p_{M|vw}}{{\rm d} v}=-2\pi{\cal K}_{M+1}(v)\,,
\eean
one finds that the minimized free energy \ref{freval2} is
\be\la{TBAcanfrengred}
 f=-{T}\log\left(1+\frac{1}{\Y_+}\right)\star \cK_1
   -\,{T}\log\left(1+\frac{1}{\Y_{M|vw}}\right)\star  \cK_{M+1}\,.
\ee
Computing the infinite sum in the second term as in appendix \ref{dervTBA} one gets
\begin{align}\la{canfreng4}
 f=&-\mu-2\gh -2c_A +J\log2\nonumber\\
 &-T\big(\log(1+{1\ov \Y_+})\star{\mathfrak s}+\log(1+{1\ov\Y_-})\star{\cal K}_2\star{\mathfrak s}
 +\log(1+\Y_{1|vw})\star{\cal K}_1\star {\mathfrak s} \big)\,,
\end{align}
where $J$ and $c_A$ are given in Table \ref{tb2} and they follow from the identities
\bea
2\pi s\star\cK_1 = \log 2\,,\quad 2\pi s\star\cK_3 =1- \log 2\,.
\eea
The free energy can be also written with dependence on the $1|w$-string 
\begin{align}\nonumber
f=&-\mu+J\log2 -T\big(\log(1+\Y_+)\star{\mathfrak s}+\log(1+\Y_-)\star{\cal K}_2\star{\mathfrak s}
 +\log(1+\Y_{1|w})\star{\cal K}_1\star {\mathfrak s} \big)\, .
\end{align}

\smallskip

Since the extended opposite A-model is the EKS model the TBA equations and free energy above describe its equilibrium states in the thermodynamic limit. 
Our canonical TBA equations look very different from  those derived in \cite{EK93},  because we used the BFFB grading for the Bethe equations while \cite{EK93} used the BBFF grading. The sets of equations nevertheless  are equivalent.  Identifying the functions $\a\,, \beta\,, \g$ used there with our Y-functions as
\bea
\a_1 = 1/\Y_+\,,\quad \beta_2 = \Y_-\,,\quad \a_{M+1}=\Y_{M|w}\,,\quad \beta_1=1/\Y_{1|vw}\,,\quad \g_{M-1}=\Y_{M|vw}\,,
\eea
we find that our simplified equations for $\Y_+$, $\Y_{M|w}\,,\ M\ge1$ and 
$\Y_{M|vw}\,,\ M\ge2$ agree with the equations (3.13)   from  \cite{EK93}, and 
our equation for $\Y_+/\Y_-$ agrees with their eq.(3.10) combined with eq.(3.8) for $n=1$. Finally the canonical equation for $\Y_{1|vw}$ follows from the combination $K_1\star\log(1+\a_1)+(\delta + K_2)\star\log\beta_1$
and  their eqs.(3.8, 3.9).\footnote{We thank Fabian Essler for matching the equations for $\Y_+/\Y_-$ and $\Y_{1|vw}$.}

Let us also mention that the  simplified equations for $\Y_-$ and $\Y_{1|vw}$  are presented in \cite{JKS97} in eq.(5.7),  
 and we match our equation for $\Y_{1|vw}$ with theirs. The equation for $\Y_-$ however in  \cite{JKS97} does not contain the crucial coupling dependent term  $4\gh/T$. It seems that to get the term from
 the quantum transfer matrix approach
 one should specify the large $v$-behaviour of $\Y_+/\Y_-$ which is determined in appendix \ref{largev}.

\subsection{Strong coupling limit}\la{exModsLargeG}

There are two natural ways to take the large $\gh$ limit of the extended model TBA equations \eqref{TBAexmod}. Taking the strict $\gh\rightarrow\infty$ limit with all other parameters fixed will lead to a half-filled model, due to the dominance of the Hubbard term favouring singly occupied sites.

Alternatively one can take $\mu\rightarrow-\infty$ simultaneously  with $\gh\rightarrow\infty$ and retain the possibility for empty sites in the Hilbert space. In particular, if one redefines the chemical potential as $\tilde \mu$ through 
\bea\la{submu}
{\tilde \mu} =\mu+2\gh\,,
\eea
the effect is that the models have been extended with the term $4\,\gh \sum_{j=1}^L \n_{j,\uparrow} \n_{j,\downarrow}$, rather than the term $4\,\gh {\mathbf V}^{\rm H}$. Then the extension does not give a cost to empty sites. Let us proceed with the discussion of this limit with the redefinition of the chemical potential and return  afterwards to the strict $\gh\rightarrow\infty$ limit. Note then that $\Y_-\rightarrow\infty$, and $\Y_{M|vw}\rightarrow\infty$ for $M\geq2$, as $\gh\rightarrow\infty$ with $\tilde \mu$ fixed. The  equations \eqref{TBAexmod} thus take the following simplified form 
\be
\begin{aligned}\notag
 \log \Y_+& = \frac{ \E_+ -\E_{1|vw}\star  {\mathfrak s}}{T}+\log\frac{1+\Y_{1|vw}}{1+\Y_{1|w}}\star  {\mathfrak s}\,,\\
\log \Y_{1|vw} &= \frac{\E_{1|vw} -2\,{\tilde \mu}}{T}+\log\Big(1+\frac{1}{\Y_{1|vw}}\Big)\star {\cal K}_{2}+ \log\Big( 1+\frac{1}{\Y_+}\Big)\star {\cal K}_1\,,\\
\log \Y_{M|w} &= I_{MN}\log\big(1+\Y_{N|w}\big)\star  {\mathfrak s} + \delta_{M1} \log\Big(1+\frac{1}{\Y_+}\Big) \star {\mathfrak s}\,,
\end{aligned}
\ee
which are the TBA equations of the supersymmetric {\it t-J} model \cite{Schlot}. The expression for free energy \eqref{TBAcanfrengred} simplifies to
\be\la{TBAcanfrengtj}
 f=-{T}\log\left(1+\frac{1}{\Y_+}\right)\star \cK_1
   -\,{T}\log\left(1+\frac{1}{\Y_{1|vw}}\right)\star  \cK_{2}\,.
\ee

Now consider the $\gh\rightarrow\infty$ limit with $\mu$ fixed. From the TBA equations \eqref{TBAexmod} and \eqref{TBAexten} note that $\Y_-\rightarrow\infty$ and $\Y_{1|vw}\rightarrow 0$, and  thus the simplified equations become
\bean
 \log \Y_+& =& \frac{ \E_+ -\E_{1|vw}\star  {\mathfrak s}}{T} - \log\big({1+\Y_{1|w}}\big)\star  {\mathfrak s}\,,\\
\log \Y_{M|w} &=& I_{MN}\log\big(1+\Y_{N|w}\big)\star  {\mathfrak s} + \delta_{M1} \log\Big( 1+\frac{1}{\Y_+}\Big) \star {\mathfrak s}\,,
\eean
where $\Y_{M|vw}$ have decoupled.
These can be simplified further by relabelling: $\Y_1=1/\Y_+$ and $\Y_M=\Y_{M-1|w}$ for $M\geq 2$.
Then the Y-functions form one set
\bean
\log \Y_{M} &=& I_{MN}\log\big(1+\Y_{N}\big)\star  {\mathfrak s} + \delta_{M1} \frac{2\pi\, J\, {\mathfrak s}}{T}\,,
\eean
and the free energy is given by
\begin{align}\la{canfrengxxx}
 f=&-\mu-2\gh -2c_A +J\log2-T\log(1+{\Y_1})\star{\mathfrak s}\,.
\end{align}
This is none other than the set of TBA equations for the XXX-spin chain \eqref{XXXTBA}. 
Comparing the values of $J$ from Table \ref{tb2} it follows that the $g\to\infty$ limit of the A-model and opposite B-model is a ferromagnetic spin chain, while
the B-model and opposite A-model  become the antiferromagnetic spin chain.

\subsection{Zero temperature phase diagram}
Here the zero temperature phase diagram of the extended B-model is obtained for the full range of coupling  $\gh\in{\mathbb R}$, see Figure \ref{phasediags}. At the end we will comment how the corresponding phase diagrams for the other three extended models can be obtained from it.

Let us break the analysis into two parts. First consider the case $\gh>0$. As in section \ref{zeroT} there are no $M|vw$- or $M|w$-strings for $M\geq2$, and as $\Y_+<\Y_-$ there are no $w$-particles at zero temperature either. Thus neither are there $z_-$-particles, as can be seen rigorously by examining the equations for zero temperature root densities \eqref{tbaygr0}. 
The TBA equations for the dressed energies are then
\be\la{T0TBABinf}
\begin{aligned}
\e_+&=\E^{\rm B_\infty}_+-2\gh-\mu-B-\e_{1|vw}\star_{Q_{1|vw}}\, {\cal K}_1\,,\\
\e_{1|vw}&=\E^{\rm B_\infty}_{1|vw}-4\gh-2 \mu - \e_+ \star_{Q_y}\, {\cal K}_1 - \e_{1|vw}\star_{Q_{1|vw}}\, {\cal K}_{2}\,,
\end{aligned}
\ee
which differ from equations  \eqref{T0TBA} only by a shift of $\mu$ by $2\gh$. Thus the phase diagram will be that of the strong coupling limit of the B-model with $\mu$ shifted as $\mu\to\mu-2\gh$. The question that remains to be addressed is how phases III and V are separated for the range $\mu \in (-2\gh,0)$. Recall that the half-filled phase is determined by the condition that $\e_{1|vw}(v)\leq 0$ for all $v\in {\mathbb R}$, and the transition to the half filled spin polarised phase is when $\e_{+}(v)\leq0$ and $\e_{1|vw}(v)=0$ for all $v\in{\mathbb R}$. From the large $v$ asymptotics of  \eqref{T0TBABinf} it follows that the phase boundary is the line $B=\mu+2+2\gh$.

Now consider $\gh<0$. Here one can no longer say that $\Y_+(v)<\Y_-(v)$ for all $v$ and so there is the possibility of $w$-particles and $z_-$-particles entering the ground state. Recall however that there can be no $w$-particles, and consequently no $z_-$-particles, unless there are $z_+$-particles. Note though that $\e_+<0$ puts restrictions on $\mu$ and $B$ that result in $\e_{1|w}>0$ and $\e_->0$. Thus equations  \eqref{T0TBABinf} govern the dressed energies for $\gh<0$ also, and again the phase diagram is that of the strong coupling limit of the B-model shifted by $\mu\to\mu-2\gh$. Let us remark  that phase V will disappear when $\gh=-1+\log 2$ and that phase IV will disappear when $\gh=-1$.

The phase diagram for the EKS model (extended opposite  A-model) is obtained from the one for the extended B-model by shifting the coupling $\gh$ above by $1$, see  \eqref{EKSvsB}. For $B=0$ our results agree with \cite{EKS}.
The corresponding phase diagrams for the extended A- and opposite  B-models are obtained by using the relationships between opposite models discussed in appendix \ref{attract}.

\section{Conclusion}\la{Conclusion}

Given the wealth of work that has been done on the Hubbard model there are many possible directions in which one can take the A- and B-models. 
To understand better the physical properties of the models it is desirable to analyse the  spectrum of excited states and the finite-size corrections at zero temperature. 
Taking into account the rather involved form of the Hamiltonians it would be interesting to consider the continuum limit of the models. 

The sets of TBA equations we obtain  depend on infinitely many functions. It would be useful to apply the fusion hierarchy approach to the thermodynamic limit and derive a finite set of nonlinear integral equations from the quantum transfer matrix  \cite{BK96,JKS98,book}.

In this paper we have restricted our focus to the parity-invariant Hubbard-Shastry models, but it might be of interest to analyse the general parity-breaking models.
Such chiral models could exhibit interesting behaviour, see for example \cite{WWZ}.

It is worth noting that there are other integrable extensions of the Hubbard model, in particular the Alcaraz-Bariev set of models \cite{AB}. These correspond to a quantum deformation of Shastry's R-matrix \cite{BKqdH}, and so provide anisotropic  deformations of the symmetries of the A- and B-models. This might allow one to introduce spin-orbit coupling to the models.  

The nonintegrable extended B-model 
interpolates  between the Hubbard,  EKS and $t$-$J$ models. Thus it might be a useful toy model for studying transitions between the various phases they exhibit. In particular the $t$-$J$ model is the established starting point for the study of the high temperature superconductivity \cite{Anderson, ZhRice}, while the Hubbard and EKS models are also mentioned in this context. Given the high symmetry of the extended B-model it might capture interesting physics that its {\it t-J} limit does not.

Another direction is to construct models for the Kondo lattice \cite{TSU,Gula}. These are models with eight states at each site, where the pseudo-vacuum carries spin-1/2 and so the site vacuum can be labelled $|\uparrow\rangle$, $|\downarrow\rangle$. The spin of the itinerant electrons can couple to the spin at the site and the models are used for the study of heavy fermion systems. Such models arise naturally from the R-matrix for the 8-dimensional atypical representation of $\sucex$ constructed in \cite{AFbound}.

Finally it is of interest to investigate how the models behave in higher dimensions. Our analysis has been largely restricted to one dimension as it is only there that the models are integrable. Higher dimensions however are more physically relevant and may also exhibit rich behaviour. Given that the models possess high symmetries there is also the  possibility that some exact statements can be made.

\section*{Acknowledgements}
We are grateful to Niklas Beisert for valuable discussions, and Fabian Essler, 
Frank G\"ohmann, Andreas Kl\"umper  
and Vladimir Korepin 
for useful comments on the 
manuscript. E.Q. thanks Max-Planck-Institut f\"ur Gravitationsphysik
Albert-Einstein-Institut for hospitality. 
The work of S.F. was supported in part by the Science Foundation Ireland under
Grant No. 09/RFP/PHY2142, and by  a
one-month Max-Planck-Institut f\"ur Gravitationsphysik
Albert-Einstein-Institut grant. The work of E.Q. was supported by the Science Foundation Ireland under
Grant No. 09/RFP/PHY2142. 


\appendix

\renewcommand{\theequation}{\thesubsection.\arabic{equation}}

\section{Appendices}

\subsection{Conventions, definitions and notations} \la{conventions}

\subsubsection*{Conventions for obtaining spin chain Hamiltonians from R-matrices}

To compute the Hamiltonians of the models we have used  the  $\sucex$-invariant R-matrix in the form given in \cite{AFbound}. Up to an overall scalar factor it is related to Shastry's R-matrix by a similarity transformation. The R-matrix depends on two independent parameters, and it can be thought of either as a function ${\mathbf R}(x^\pm_1,x^\pm_2)$ of the four parameters $x^\pm_1\,, x^\pm_2$ subject to the two constraints 
\be\la{xpmconstr2}
x_k^+ + \frac{1}{x_k^+} - x_k^- -\frac{1}{x_k^-}= 4\, i\,\uh\,,\quad k=1,2\,,
\ee
or as a function ${\mathbf R}(z_1,z_2)$ where $z_1\,,z_2$ parametrize the tori  defined by \eqref{xpmconstr2}.
One can show that $x^\pm_1\,, x^\pm_2$ and ${\mathbf R}(z_1,z_2)$ are meromorphic functions of $z_1$ and $z_2$ \cite{AFbound}. For this reason it is safer to use the $z$-parametrization for computations. The Hamiltonian is given by ${\mathbf H} = \sum_{j=1}^L {\mathbf H}_{j,j+1}$ where ${\mathbf H}_{j,j+1}$ is computed as 
\be\la{scham2}
{\mathbf H}_{12}= i\, {\cal N}_{\mathbf H}(z)\, {\mathbf P}_{12} \partial_1 {\mathbf R}_{12}(z_1,z)|_{z_1=z}\,,
\ee
where the normalization constant ${\cal N}_{\mathbf H}(z)$ is chosen so that the dispersion relation would be given by \eqref{disprels}, and since the spin chain is homogeneous it is sufficient to compute only ${\mathbf H}_{12}$ . We find the following normalization constants
\begin{equation}
\begin{array}{ccc}
({\rm H}) &~~~~z=\omega_1/2 -\om_2/2\,,& {\cal N}_{{\mathbf H}^{\rm H}}(z)= 2\uh\,,~~~~~~\\
({\rm A}) &z= -\om_2/2\,,~~~&~~~~{\cal N}_{{\mathbf H}^{\rm A}}(z)= 2\uh\sqrt{1+\uh^2}\,,\\
({\rm B}) &z=\omega_1/2 \,,~~~~~&~~~~~~ {\cal N}_{{\mathbf H}^{\rm B}}(z)= 2\uh/\sqrt{1+\uh^2}\,.
\end{array}\nonumber
\end{equation}
Here $\om_1=2K(-1/\uh^2)$ and $\om_2=2iK(1+1/\uh^2)-2K(-1/\uh^2)$ are the real and imaginary half-periods of the torus defined by \eqref{xpmconstr2}.

One can also obtain the same Hamiltonians  
as
\be\la{scham3}
{\mathbf H}_{12}=  i\,  {\cal N}_{\mathbf H}(x^\pm)\,  {\mathbf P}_{12} {\partial\ov\partial{x_1^+}} {\mathbf R}_{12}(x^\pm_1,x^\pm)|_{x^\pm_1=x^\pm}\,,
\ee
where $x_1^-$ is considered as a function of $x_1^+$. One then gets 
\be
{\cal N}_{\mathbf H}(x^\pm) =i\, {x^+\ov x^-}(x^+-x^-)\sqrt{1-x^-{}^2\ov 1-x^+{}^2}\,,
\ee
and therefore
\begin{equation}
\begin{array}{ccc}
({\rm H}) &x^+=1/x^-=\infty, ~~~~~~~~~~~~~~~~~~& {\cal N}_{{\mathbf H}^{\rm H}}(x^\pm)\to x^+/x^-\,,\\
({\rm A}) &x^+=-1/x^-=i(\uh+\sqrt{1+\uh^2}),  &~~~~{\cal N}_{{\mathbf H}^{\rm A}}(x^\pm)= 2\uh\sqrt{1+\uh^2}\,,\\
({\rm B}) &x^+=-x^-=i(\uh+\sqrt{1+\uh^2}), ~~&~~~ {\cal N}_{{\mathbf H}^{\rm B}}(x^\pm)= 2\sqrt{1+\uh^2}\,.
\end{array}\nonumber
\end{equation}
The resulting ${\mathbf H}_{12}$ is represented either as a $16\times 16$ matrix or as a differential operator and can be rewritten in terms of the fermion creation and annihilation operators $\cd_{k,\uparrow}\,,  \cm_{k,\uparrow}\,, \cd_{k,\downarrow} \,, \cm_{k,\downarrow}$  by using the identification 
\be
\cd_{\uparrow} \equiv E_{31}-E_{24}
 \, ,\quad
 \cm_{\uparrow} \equiv E_{13}-E_{42}
\, ,\quad  \cd_{\downarrow} \equiv E_{41}+E_{23}
\, ,\quad  \cm_{\downarrow}  \equiv E_{14}+E_{32}
\,.
 \ee
\be\nonumber
 \cd_{\uparrow}\otimes I_4\ \leftrightarrow \
 \cd_{1,\uparrow} \, ,\quad
 \cm_{\uparrow}\otimes I_4\ \leftrightarrow \
 \cm_{1,\uparrow} \, ,\quad  \cd_{\downarrow}\otimes I_4\ \leftrightarrow \
 \cd_{1,\downarrow} \, ,\quad  \cm_{\downarrow} \otimes I_4\ \leftrightarrow \
 \cm_{1,\downarrow}  \,.
 \ee
 \be\nonumber
I^g\cdot I_4\otimes \cd_{\uparrow}\cdot I^g\ \leftrightarrow \
 \cd_{2,\uparrow} \, ,\ \
I^g\cdot  I_4\otimes \cm_{\uparrow}\cdot  I^g\ \leftrightarrow \
 \cm_{2,\uparrow} \, ,\ \  I^g\cdot  I_4\otimes \cd_{\downarrow}\cdot  I^g\ \leftrightarrow \
 \cd_{2,\downarrow} \, ,\ \   I^g\cdot  I_4\otimes \cm_{\downarrow}\cdot  I^g\ \leftrightarrow \
 \cm_{2,\downarrow}  \,.
 \ee
Here $E_{ij}$ are $4\times 4$ matrix unities, $I_4$ is a $4\times 4$ identity matrix, and $I^g = \sum_{i,j}(-1)^{\e_i\e_j}E_{ii}\otimes E_{jj}$ is the graded identity where $\e_1=\e_2=0$, $\e_3=\e_4=1$.

Finally, to get the kinetic term in the canonical form \eqref{THub} one should perform the canonical transformation \eqref{Ual} with $\a = -\pi/2$ for the Hubbard model, $\a = \pi$ for the A-model, and $\a = -\pi/2$ for the B-model.

\subsubsection*{Matching the notations and conventions}

Most of 
our notations and conventions come from   \cite{AFS09}, and 
here we compare them to  those of \cite{book}.

In the Bethe ansatz we denote particles momenta as $p_j$ and auxiliary roots as $w_j$, so they are related to the ones in \cite{book} as $p_j \leftrightarrow k_j$, $w_j \leftrightarrow \Lambda_j$.

In the string hypothesis a $M|w$-string is a $\Lambda$ string of length $M$,
a $w$-particle is a $\Lambda$-string of length 1, 
a $M|vw$-string is a $k$-$\Lambda$ string of length $M$, and
$y$-particles could have been called $k$-particles.

In the TBA equations the Y-functions are related to the ones in \cite{book} as $Y_{M|w} \leftrightarrow \eta_M$, $Y_{M|vw}  \leftrightarrow \eta_M'$, $Y_{-}(\sin(k))  \leftrightarrow \zeta(k)\,,\ |k|\le\pi/2$ and $Y_{+}(\sin(k))  \leftrightarrow \zeta(k)\,,\ |k|\ge\pi/2$.

\subsubsection*{Kernels and S-matrices}

We use several types of convolutions defined in (\ref{star}--\ref{star3}) and \eqref{stara}.
The TBA equations  
involve convolutions with three types of kernels
which we list below
\begin{alignat}{2}
s (v) & = \frac{1}{2 \pi i} \, \frac{d}{dv} \log S(v)= {1 \ov 4\uh \cosh {\pi  v \ov 2\uh }}\,,\quad S(v)=-\tanh\big( \frac{\pi v}{4\uh} -\frac{i\pi}{4}\big)\,,
\la{skern}\\
K_M (v) &= \frac{1}{2\pi i} \, \frac{d}{dv} \, \log S_M(v) = \frac{1}{\pi} \, \frac{\uh\, M}{ v^2+\uh^2M^2}\,,\quad S_M(v)= \frac{v - i\,\uh\, M}{v + i\,\uh\, M} \,, \la{KQkern}\\
K_{MN}(v) &= \frac{1}{2\pi i} \, \frac{d}{dv} \, \log S_{MN}(v)=K_{M+N}(v)+K_{N-M}(v)+2\sum_{j=1}^{M-1}K_{N-M+2j}(v)\,,\la{KMNkern}\\
S_{MN}(v) &=S_{M+N}(v)S_{N-M}(v)\prod_{j=1}^{M-1}S_{N-M+2j}(v)^2 =S_{NM}(v)\,.\la{SMNkern}
\end{alignat}
Note that the kernels satisfy the following identities
\be\la{kerKid}\begin{aligned}
&1\star s = {1\ov 2}\,,\quad 1\star K_M = 1\,,\quad K_M\star K_N = K_{M+N}\,,\\
&K_1-s\star K_2 = s\,,\quad K_{M+1}-s\star K_M-s\star K_{M+2} =0\,.
\end{aligned}
\ee
\subsection{Deriving the TBA equations} \la{dervTBA}
\subsubsection*{String hypothesis}

The string hypothesis for the models has been formulated in section \ref{stringhyp}, and according to it in the thermodynamic limit 
every solution of the Bethe equations \eqref{BE1} corresponds to a particular configuration of Bethe strings and consists of
\begin{enumerate}
\item $N_{+}$ $ y_+$-particles with rapidities $v_{+,k}$
\item $N_{-}$ $ y_-$-particles with rapidities $v_{-,k}$
\item $N_{M|{vw}}$ $ M|{vw}$-strings, $M=1,2,\ldots,\infty$ with rapidities $v_{M,k}$
\item $N_{M|{w}} $ $M|{w}$-strings, $M=1,2,\ldots,\infty$ with rapidities $w_{M,k}$.
\end{enumerate}
The numbers $N_{+}$, $N_{-}$, $N_{M|{vw}}$, $N_{M|{w}}$ are called the occupation numbers of the root configuration under consideration and they are related to $N, M$ as follows
\bean
N = N_{+}+N_{-}+\sum_{Q=1}^{\infty}2 \,Q\, N_{Q|{vw}},\quad
M = \sum_{Q=1}^{\infty} Q \,N_{Q|{vw}}+\sum_{Q=1}^{\infty} Q\,N_{Q|{w}}.
\eean

\subsubsection*{Bethe equations for string configurations}

Next we rewrite  the Bethe equations in terms
of the string configurations. The S-matrices presented in equations (\ref{KQkern},\ref{SMNkern}) are used. 
The Bethe equation for a $y_{\pm}$ particle with rapidity $v_{\pm,k}$ is
\bea\la{BEypm}
1= e^{i\, p_{\pm}(v_{\pm,k})\, L} 
 \prod_{M=1}^{\infty} \prod_{i=1}^{N_{M|vw}}S_M(v_{\pm,k}-v_{M,i}) \,
  \prod_{N=1}^{\infty} \prod_{j=1}^{N_{N|w}}S_M(v_{\pm,k}-w_{N,j}) \,,
\eea
where the $p_{\pm}$ are given by equations (\ref{pk}) and \eqref{ypm}. 
The Bethe equation for an $M|vw$-string with root $v_{M,k}$ is 
\bea
(-1)^M = e^{i\, p_{M|vw}(v_{M,k}) \,L}
\prod_{i=1}^{N_{+}}S_M(v_{M,k} -v_{+,i})
\prod_{j=1}^{N_{-}}S_M(v_{M,k} -v_{-,j})
 \prod_{N=1}^{\infty} \prod_{l=1}^{N_{N|vw}}S_{MN}(v_{M,k}-v_{N,l}),
\eea
and the momentum $p_{M|vw}$ is given by \eqref{pEvw}.
Finally the Bethe equation for an $M|w$-string with root $w_{M,k}$ is
\bea\la{BEw}
 (-1)^M \prod_ {N=1}^{\infty} \prod_{l=1}^{N_{N|w}}S_{MN}(w_{M,k}-w_{N,l}) = \prod_{i=1}^{N_{+}}S_M(w_{M,k} -v_{+,i})
\prod_{j=1}^{N_{-}}S_M(w_{M,k} -v_{-,j})
\,.
\eea

\subsubsection*{TBA equations for  densities}

We introduce densities  $\rho(v)$ of particles, and densities $\bar{\rho}(v)$ of holes in the standard way
and obtain the following TBA equations for the densities 
 by taking the log derivatives of the Bethe equations for the string configurations
\bea\la{tbaygr}
\rho_{+} + {\bar{\rho}}_{+} &=&-\frac{1}{2 \pi}\frac{d  p_+}{d v} - \sum_{M=1}^\infty K_M\star\left(\rho_{M|vw}+\rho_{M|w}\right)  \,,~~~~\\
\la{tbaylss}
\rho_{-} + {\bar{\rho}}_{-} &=&\frac{1}{2 \pi}\frac{d p_-}{d v}+ \sum_{M=1}^\infty K_M \star\left(\rho_{M|vw}+\rho_{M|w} \right)  \,,~~~~
\\\la{tbavw}
\rho_{M|vw} + \br_{M|vw} &=&-\frac{1}{2 \pi}\frac{d  p_{M|vw}}{d v}- \sum_{N=1}^\infty K_{MN}\star \rho_{N|vw} 
-K_M\,\bar\star\,\left(\rho_{+} +\rho_{-} \right),~~~~~~~~~~
\\\la{tbaw}
\rho_{M|w} + \br_{M|w} &=&-\sum_{N=1}^\infty K_{MN}\star\rho_{N|w} +  K_M\,\bar\star\,\left(\rho_{+}
 +\rho_{-} \right)   \,.~~~~
\eea
 Here we have taken into account that  for all the models $\frac{d p_+}{d v}<0$ and $\frac{d p_-}{d v}>0$, and  $\frac{d p_{M|vw}}{d v}<0$. 
The convolution $\bar\star$ is defined as 
\bea\la{stara}
g\bstar h &=& g\hstar h\quad {\rm for\ the\ Hubbard\ and\ A-models}\,,\\\nonumber
g\,\bar\star\,  h &=& g\cstar h\quad {\rm for\ the \ B-model}\,,
\eea
and the kernels are defined by \eqref{KQkern} and \eqref{KMNkern}. It is worth noting that 
\be\la{denph}
1\bstar(\rho_{+} +\rho_{-} + {\bar{\rho}}_{+}+ {\bar{\rho}}_{-})=1\,.
\ee

\medskip

Let us briefly consider the zero temperature limit of the equations for densities for A- and B-models. 
As is discussed in section \ref{zeroT} in this limit  $Y_{M|vw}>1$ for $M\ge 2$ and $Y_{M|w}>1$ for $M\ge 1$, and therefore the corresponding densities of particles go to zero. 
For the remaining string types there are intervals $Q_k$ where 
the densities of holes vanish, and outside which
the densities of particles vanish.
Then the equations above simplify 
\begin{align}\nonumber
\rho_{+}=&-\frac{1}{2 \pi}\frac{d  p_+}{d v} - K_1\star_{Q_{1|vw}}\rho_{1|vw} \,,\quad v\in Q_+\,,~~~~\\\la{tbaygr0}
\rho_{-} =&\frac{1}{2 \pi}\frac{d p_-}{d v}+ K_1 \star_{Q_{1|vw}}\rho_{1|vw} \,,\quad v\in Q_-\,,~~~~
\\\nonumber
\rho_{1|vw} =&-\frac{1}{2 \pi}\frac{d  p_{1|vw}}{d v}- K_{2}\star_{Q_{1|vw}} \rho_{1|vw} 
-K_1\star_{Q_{+}}\rho_{+} - K_1\star_{Q_{-}}\rho_{-} \,,\quad v\in Q_{1|vw}\,.~~~
\end{align}

\subsubsection*{Canonical TBA equations}
The free energy in terms of the string densities takes the following form
\be
\begin{aligned}\la{freval}
f(\mu,B,T)=&\int {\rm d}v (\E_+-\mu-B)\rho_{+}+\int {\rm d}v (\E_- -\mu-B)\rho_{-}\\
&+\sum_{M=1}^{\infty}\int {\rm d}v (\E_{M|vw}- 2 M\, \mu)\rho_{M|vw}
+\sum_{M=1}^{\infty}\int {\rm d}v\, 2 M \,B \,\rho_{M|w}-T\,s.
\end{aligned}
\ee
Then the entropy per site is given by
\bea\nonumber
s = \int {\rm d}v\,\left(
 \cs\left(\rho_{+} \right)+\cs\left(\rho_{-} \right) + \sum_{M=1}^\infty\left( \cs\big(\rho_{M|vw} \big)+\cs\big(\rho_{M|w} \big) \right) \right)\,,
\eea
 where $\cs(\rho)$ denotes the entropy function of densities of particles and holes
\bea
\cs(\rho)= \rho\log\left(1+{\br\ov\rho}\right) + \br\log\left(1+{\rho\ov\br}\right)\,.
\eea

The TBA equations are obtained by minimizing the free energy and introducing 
the Y-functions 
\bea
Y_{\pm} ={\bar{\rho}_\pm\ov \rho_\pm}\,,\quad Y_{M|vw} ={\bar{\rho}_{M|vw} \ov \rho_{M|vw} }\,,\quad Y_{M|w} ={\bar{\rho}_{M|w} \ov \rho_{M|w} }\,,
\eea
they
can be written in the form  
\begin{align}
\log Y_{\pm} &= \frac{\E_\pm-\mu-B}{T}+\log \frac{1+\frac{1}{Y_{N|vw}}}{1+\frac{1}{Y_{N|w}}}\star K_{N},~~~~~\la{TBAy}\\
\log Y_{M|vw} &= \frac{\E_{M|vw}-2\,M\,\mu}{T}+\log\left(1+\frac{1}{Y_{N|vw}}\right)\star K_{NM}-\log\left(1+\frac{1}{Y} \right)\circledast K_M\,,\la{TBAvw}\\
\log Y_{M|w} &= \frac{2\,M\,B}{T}+\log\left(1+\frac{1}{Y_{N|w}}\right)\star K_{NM}-\log\left(1+\frac{1}{Y} \right)\circledast K_M\,,~~~~~\la{TBAw}
\end{align}
where we sum over $N=1,2,\ldots$ and  recall that $Y_\pm(v) \equiv Y(v\pm i0)$. 
We refer to the TBA equations written in this form as the {\it canonical} ones.
The minimized free energy is 
\be\la{TBAcanfreng}
 f=-{T\ov 2\pi}\log\left(1+\frac{1}{Y}\right)\circledast \frac{{\rm d}\,p}{{\rm d}v}
   +\,{T\ov 2\pi}\log\left(1+\frac{1}{Y_{M|vw}}\right)\star  \frac{{\rm d}\,p_{M|vw}}{{\rm d}v}\,.
\ee
The infinite sum in the second term can be simplified as will be discussed shortly.  

\subsubsection*{Simplifying the TBA equations}
The above equations can be rewritten in a simpler "local" form. To achieve this we introduce the following 
kernel
 \bea\la{invK0} \left( K  +  \delta\right)_{MN}^{-1} (v)\equiv \delta_{MN}\delta(v)
- I_{MN}\, s(v)
\eea 
with $ I_{MN}=\delta_{M+1,N}+ \delta_{M-1,N}\,,\ M\ge2 $ and $ I_{1N}=\delta_{2N}$,
that is inverse to $ K_{NQ}  + \delta_{NQ}$:
 \bea \la{invKK}
\sum_{N=1}^\infty\left( K +  \delta\right)_{MN}^{-1}\star \left( K_{NQ}  +
\delta_{NQ}\right)= \sum_{N=1}^\infty\left( K_{QN}  +
\delta_{QN}\right)\star \left( K +  \delta\right)_{NM}^{-1} =\delta_{MQ}\,,~~~~~ \eea
which can be also written in the following convenient form
 \bea \la{invKKb}
\sum_{N=1}^\infty\left( K +  \delta\right)_{MN}^{-1}\star K_{NQ} = \sum_{N=1}^\infty\,K_{QN}\star \left( K +  \delta\right)_{NM}^{-1} =I_{MQ}\,s\,.~~~~~ \eea
Here and in what follows we often use the convention $\delta_{MN}(v)\equiv \delta_{MN}\delta(v)$ when this will not cause confusion.
This obeys the useful identities
\bea\la{kmnikm}
&&\sum_{N=1}^\infty\left( K +  \delta\right)_{MN}^{-1} \star K_{N} =\sum_{N=1}^\infty K_{N} \star\left( K +  \delta\right)_{NM}^{-1}=\delta_{M1}\, s\,,
\\
\la{evwki}
 &&\E_{N|vw}\star (K+ \delta)_{NM}^{-1}= -\delta_{1M}\,\E\circledast  s =\delta_{1M}\left(\E_+ -\E_-\right)\bstar s\,.
\eea

Now we can proceed with simplifying the canonical TBA equations. Subtracting the TBA equations \eqref{TBAy} for $y_\pm$-particles gives
\bea
 \log\frac{Y_+}{Y_-}&=&\frac{\E_+ - \E_-}{T}\,. ~~~~\la{Yyym}
\eea
Then applying the inverse kernel to the equations for the $vw$- and $w$-strings and using the identities above and \eqref{Yyym} gives the simplified TBA equations:
\bea\la{ymvweqn}
\log Y_{M|vw} &=&I_{MN}\log\left(1+Y_{N|vw}\right)\star s -\delta_{M1}\log\left(1+{Y} \right)\circledast s\,,\qquad\\
\log Y_{M|w} &=& I_{MN}\log\left(1+Y_{N|w}\right)\star s -\delta_{M1} \log\left(1+\frac{1}{Y} \right)\circledast s\,,
\eea
where we took into account that $(K+ \delta)^{-1}$ annihilates the $\mu$ and $B$ dependent terms.

The simplified TBA equations can be used to compute the infinite sums appearing in the canonical equations for $y$-particles. We show how to do this on the example of $vw$-strings. Since a similar sum appears in expression \eqref{TBAcanfreng} for the free energy we do the computation  by assuming that we are given a kernel $\mK_M$ satisfying the identity\footnote{ This method was applied in \cite{AFS09} to more general kernels but under an assumption of $Y_M$-functions approaching 1 at large $M$.}
\bea\la{kmnikm2}
\sum_{N=1}^\infty\left( K +  \delta\right)_{MN}^{-1} \star\mK_{N} =\mK_{M} -I_{MN}s\star \mK_{N}=\delta_{M1}\, \delta \mK\,,
\eea
where $\delta \mK$ is any kernel. We want to compute the following sum
\bea
\sum_{M=1}^\infty\log\left(1+{1\ov Y_{M|vw}}\right)\star \mK_{M} \,.
\eea
To this end \eqref{ymvweqn} is rewriten in the form
\bean
\log Y_{M|vw}^{reg} -I_{MN}\log Y_{N|vw}^{reg}\star s=I_{MN}\log\big(1+{1\ov Y_{N|vw}}\big)\star s -\delta_{M1}\Big({\E\ov T}+\log\big(1+{1\ov Y} \big)\Big)\circledast s\,,~~~
\eean
where on the l.h.s. of this equation we replaced $Y_{M|vw}$ by 
$Y_{M|vw}^{reg} = Y_{M|vw} e^{2M\mu/T}$ to make sure it asymptotes to 1 at large $M$. This does not change the equation because the kernel $s$ integrates to $1/2$, that is $1\star s = 1/2$. This replacement is necessary because at the next step we multiply this equation by $\mK_{M}$, take the sum over $M$ and use the identities \eqref{kmnikm2} and \eqref{evwki} to  get
\bea
\log Y_{1|vw}^{reg}\star \delta \mK&=&\log\left(1+{1\ov Y_{M|vw}}\right)\star \mK_{M} -\log\left(1+{1\ov Y_{1|vw}}\right)\star \delta \mK \\\nonumber
&+& {\E_{1|vw}\star \delta \mK\ov T}-\log\left(1+{1\ov Y} \right)\circledast s\star \mK_1\,.~~~
\eea
From this equation and the definition of $Y_{M|vw}^{reg} $ we immediately obtain
\begin{align}\nonumber
\log\left(1+{1\ov Y_{M|vw}}\right)\star \mK_{M} =\log\left(1+Y_{1|vw}\right)\star \delta\mK&- \frac{\E_{1|vw}-2\mu }{T}\star \delta\mK \\
&+\log\left(1+{1\ov Y} \right)\circledast s\star \mK_1\,.~~~
\la{sumYvw}
\end{align}
A similar formula can be derived for the infinite sum with $Y_{M|w}$-functions
 \bean
\log\left(1+{1\ov Y_{M|w}}\right)\star \mK_{M} =\log\left(1+Y_{1|w}\right)\star \delta\mK- \frac{2B}{T}\star \delta\mK +\log\left(1+{1\ov Y} \right)\circledast s\star \mK_1\,.~~~
\eean

In our case $\mK_M = K_M$, $\delta\mK = s$, and
subtracting these two equations one obtains 
\bean
\log\frac{1+\frac{1}{Y_{M|vw}}}{1+\frac{1}{Y_{M|w}}}\star K_{M}=\log\frac{1+Y_{1|vw}}{1+Y_{1|w}}\star s +\frac{\mu +B}{T}
- \frac{\E_{1|vw}}{T}\star s\,.
\eean
The infinite sum is substituted for in \eqref{TBAy}
to give the simplified equation for $Y_+Y_-$
\bean
 \log Y_+ Y_- & =& \frac{\E_+ +\E_- -2\E_{1|vw}\star s}{T}+2\log\frac{1+Y_{1|vw}}{1+Y_{1|w}}\star s\,.
 \eean
 Combining the equations for $Y_+/Y_-$ and $Y_+Y_-$ gives
 \bea\la{simTBAY}
 \log Y& =& \frac{\E-\E_{1|vw}\star s}{T}+\log\frac{1+Y_{1|vw}}{1+Y_{1|w}}\star s\,.
 \eea

 \subsubsection*{Simplifying the free energy}
 
 Eq.\eqref{sumYvw} can be also used to compute the sum appearing in the free energy expression \eqref{TBAcanfreng}. 
First we rewrite  \eqref{sumYvw} as
\begin{align}\nonumber
&\log\left(1+{1\ov Y_{M|vw}}\right)\star \mK_{M} =\log\left(1+Y_{1|vw}\right)\star \delta\mK+ \frac{2\mu }{T}\star \delta\mK +\log\left(1+{Y} \right)\circledast s\star \mK_1\,.~~~\\
\la{sumYvw2}
\end{align} 
Then we notice that 
 \bean
(K+1)_{MN}^{-1}\star\frac{{\rm d}\,p_{N|vw}}{{\rm d}v}&=&-\delta_{M1}s\circledast \frac{{\rm d}\,p}{{\rm d}v} =\delta_{M1}s\bstar\left(\frac{{\rm d}\,p_+}{{\rm d}v} - \frac{{\rm d}\,p_-}{{\rm d}v} \right)\,,
\eean
and therefore in this case $\mK_M =  \frac{{\rm d}\,p_{M|vw}}{{\rm d}v}$, 
$\delta\mK = -s\circledast \frac{{\rm d}\,p}{{\rm d}v} $. Thus eq.\eqref{sumYvw2} becomes
\begin{align}\nonumber
\log\left(1+{1\ov Y_{M|vw}}\right)\star \frac{{\rm d}\,p_{M|vw}}{{\rm d}v}=&-\log\left(1+Y_{1|vw}\right)\star s\circledast \frac{{\rm d}\,p}{{\rm d}v} - \frac{2\mu }{T}\star s\circledast \frac{{\rm d}\,p}{{\rm d}v} \\
& +\log\left(1+{Y} \right)\circledast s\star  \frac{{\rm d}\,p_{1|vw}}{{\rm d}v}\,.~~~
\la{sumYvw3}
\end{align}
Taking into account that
$
1\star s \bstar\big(\frac{{\rm d}\,p_+}{{\rm d}v}-\frac{{\rm d}\,p_-}{{\rm d}v}\big) = - \pi\,,
$
one gets 
\begin{align}\la{canfreng2}
 f=-\mu&-{T\ov 2\pi}\log\Big(1+{1\ov Y}\Big)\circledast  \frac{{\rm d}\,p}{{\rm d}v} +{T\ov 2\pi}\log\left(1+{Y}\right)\circledast s\star  \frac{{\rm d}\,p_{1|vw}}{{\rm d}v}
 \\ \nonumber &-\,{T\ov 2\pi}\log\left(1+Y_{1|vw}\right)\star s\circledast \frac{{\rm d}\,p}{{\rm d}v}\,.
\end{align}
Finally using eq.(\ref{simTBAY}) and identity \eqref{Edp} the free energy can be rewritten as
\begin{align}\la{canfreng3}
 f=-\mu&+{1\ov 2\pi}\E_{1|vw}\star s\circledast   \frac{{\rm d}\,p}{{\rm d}v}-{T\ov 2\pi}\log\left(1+{1\ov Y}\right)\circledast \Big( \frac{{\rm d}\,p}{{\rm d}v} - s\star  \frac{{\rm d}\,p_{1|vw}}{{\rm d}v}\Big)
 \\ \nonumber &-\,{T\ov 2\pi}\log\left(1+Y_{1|vw}\right)\star s\circledast \frac{{\rm d}\,p}{{\rm d}v}\,,
\end{align}
or with dependence on the $1|w$-string in the from \eqref{freng}.

\subsection{Attractive vs. repulsive models}\la{attract}

For any given model one can consider ${\mathbf H}$ or $\tH$ as the Hamiltonian. Here it is useful to distinguish the two cases by the sign of the coefficient of the 
$\sum_j\n_{j\uparrow}\n_{j\downarrow}$ term, calling the model repulsive if the coefficient is positive and attractive if the coefficient is negative. Throughout the text the repulsive case was singled out as it appears to be the more physically relevant. A primary application of the models presented is to systems of spin-$\frac{1}{2}$ electrons and these are driven by the Coulomb repulsion. Nevertheless, it may also be of interest to examine the attractive cases.

The TBA equations for the attractive Hubbard model were derived in 
\cite{LS}. Here we show how the TBA equations and free energy for the attractive Hubbard-Shastry models can be straightforwardly obtained from those for the repulsive models. 
Up to a unitary transformation, see \eqref{opHam}, an attractive model differs from the repulsive one by the sign of the Hamiltonian. They share the same set of Bethe equations but have opposite signs of the dispersion relations. 
 Thus the only change to the thermodynamic Bethe ansatz analysis is the sign of the dispersion relation and therefore  the TBA equations for the attractive models are
\bea\la{tstbay}
 \log \tY& =& -\frac{\E -\E_{1|vw}\star s}{T}+\log\frac{1+\tY_{1|vw}}{1+\tY_{1|w}}\star s\,,\\ \la{tstbayvw}
\log \tY_{M|vw}(v) &=& I_{MN}\log\left(1+\tY_{N|vw}\right)\star s -\delta_{M1} \log(1+\tY){\circledast} s\,,\\\la{tstbayw}
\log \tY_{M|w}(v) &=& I_{MN}\log\left(1+\tY_{N|w}\right)\star s -\delta_{M1} \log (1+\frac{1}{\tY}){\circledast} s\,,
\eea
and they  are supplemented with the large $M$ asymptotics 
\bea
\lim_{M\rightarrow\infty} \frac{\log \tY_{M|vw}}{M}=-\frac{2\,\tmu}{T}\,,\qquad 
\lim_{M\rightarrow\infty} \frac{\log \tY_{M|w}}{M}=\frac{2\,\tB}{T}\,.\la{tMasym}
\eea
Here and in what follows tilded quantities refer to the attractive models, and untilded ones to the corresponding repulsive ones. The free energy of the attractive models are given by either \eqref{canfreng3} or  \eqref{freng} with the replacement $\E\to -\E$,  $\E_{1|vw}\to -\E_{1|vw}$, $Y\to \tY$,  $Y_{1|w}\to\tY_{1|w}$ and $Y_{1|vw}\to\tY_{1|vw}$. 

The TBA equations  are identical to the ones for the  repulsive models (\ref{stbay} - \ref{stbayw}) under the identification
\bea
\tY&=&\frac{1}{Y},\quad
\tY_{M|vw}=Y_{M|w}\,,\quad
\tY_{M|w}=Y_{M|vw}\,,\quad \tmu=-B\,,\quad \tB=-\mu\,.
\eea
By using this identification one can relate the free energy of the attractive models to those of the repulsive one. Taking into account the identities
\be\la{Edp}
\E\circledast s\star  \frac{{\rm d}\,p_{1|vw}}{{\rm d}v}=\E_{1|vw}\star s{\circledast}\frac{{\rm d}\,p}{{\rm d}v} \,,\quad \E {\circledast}\frac{{\rm d}\,p}{{\rm d}v} =\E_{\pi/2}
\ee
where $\E_{\pi/2}$ is an electron energy at $p=\pi/2$, that is $\E_{\pi/2}=-2\uh$ for the Hubbard model, $\E_{\pi/2}=-2\sqrt{1+\uh^2}$ for the A-model and $\E_{\pi/2}=0$ for the B-model, one finds
\be\la{tff}
\tf(\tmu,\tB,T)=f_{\{\mu\to-\tB,B\to-\tmu\}}-\tmu-\tB - \E_{\pi/2}\,.
\ee
Since  $\E_{\pi/2}$ depends only on $\uh$  this will not affect the physical properties as they are related to derivatives of $f$. In fact  $\E_{\pi/2}$ can be removed by redefining the Hubbard interaction potential \eqref{UHub} as follows
 \be
  {\mathbf V}^H_{j,k} \to {\mathbf V}^H_{j,k} +{1\ov4}= {1\ov 2}(\n_{j,\uparrow}-{1\ov2})(\n_{j,\downarrow}-{1\ov2})+{1\ov 2}(\n_{k,\uparrow}-{1\ov2})(\n_{k,\downarrow}-{1\ov2})\,,
\ee
which is the potential used in \cite{book}. 

The same relation between the free energies of the attractive and repulsive models can be also found by using the  Woynarovich transformation \cite{W} which is a superposition of the parity transformation and the partial particle-hole transformation of the spin-up electrons, see \cite{book} for detail. Under the   transformation the Hubbard potential \eqref{UHub} transforms as ${\mathbf V}^H_{j,k} \to -{\mathbf V}^H_{j,k} -\frac{1}{2}$, and all the other kinetic and potential terms just change their signs. The relation  \eqref{tff} then follows from the definition \eqref{deffreng} for free energy and that 
${\mathbf N} \to -2{\mathbf S}^z+1$ and ${\mathbf S}^z\to -{1\ov2} {\mathbf N} +{1\ov2} $ under the  Woynarovich transformation.

\subsection{Algebraic limit of TBA equations}\la{largev}

As was discussed in section \ref{weaklimit}, in the weak coupling limit and for  values of $v$ on $\I^{\rm A}$ and $\I^{\rm B}$ the TBA equations of the Hubbard and B-models reduce to the  set of algebraic equations (\ref{u0TBAp}-\ref{u0TBAw}).
The general solution to these  equations is well-known, see e.g. \cite{Takbook}, and can be written in the form
\bea\notag
1+Y_{M|vw} &=& \left({\sinh(f_{vw}+M)c_{vw}\ov \sinh c_{vw}}\right)^2 \,,\quad 1+Y_{M|w} = \left({\sinh(f_{w}+M)c_{w}\ov \sinh c_{w}}\right)^2 \,,
 \\
\notag
\frac{1+Y_+}{1+Y_-}\ &=& \left({\sinh f_{vw}c_{vw}\ov \sinh c_{vw}}\right)^2 \,, ~~~~~~~~~~~\quad 
\frac{1+\frac{1}{Y_+}}{1+\frac{1}{Y_-}} = \left({\sinh f_{w}c_{w} \ov \sinh c_{w}}\right)^2 \,.
\eea
Here  the constants $c_{vw}$ and $c_w$ are fixed by the large $M$ asymptotics \eqref{Masym} of Y-functions
$
c_{vw}= -{\mu\ov T}=\bc \,,\ c_{w}={B\ov T}=\bs \,,$
and the two remaining functions are then found from equations (\ref{u0TBAp})
\bea\nonumber
\sinh f_{vw}\bc&=&\frac{e^{\bd } \sinh \bc \left(\cosh \bc+e^{-\bd } \cosh
   \bs\right)}{\sqrt{2 \cosh \bc \cosh \bs \cosh \bd
   +\cosh ^2\bc+\cosh ^2\bs+\sinh ^2\bd }}\,,\\\nonumber
\sinh f_{w}\bs&=&\frac{\sinh \bs \left(\cosh \bc+e^{-\bd } \cosh
   \bs\right)}{\sqrt{2 \cosh \bc \cosh \bs \cosh \bd
   +\cosh ^2\bc+\cosh ^2\bs+\sinh^2 \bd}}\,,
\eea
with $\beta_\Delta=\frac{\Delta}{T}$.
The explicit solution to the above TBA equations is then
{\small
\begin{align}
\nonumber
1+Y_{M|vw} &= \frac{e^{2\bd } \text{csch}^2\bc \left(2 e^{-\bd } \cosh \bs \sinh (M\bc+\bc)+e^{-2\bd } \sinh M \bc
  +\sinh (M\bc+2\bc)\right)^2}{4 \left(2 \cosh \bc \cosh
   \bs \cosh \bd +\cosh ^2\bc+\cosh ^2\bs+\sinh
   ^2 \bd \right)} \,,
 \\
\nonumber
1+Y_{M|w} &=\frac{e^{2\bd } \text{csch}^2\bs \left(2 e^{-\bd } \cosh
   \bc \sinh (M \bs +\bs)+e^{-2\bd } \sinh ( M\bs+2\bs)+\sinh M \bs  \right)^2}{4 \left(2 \cosh \bc \cosh
   \bs \cosh \bd+\cosh ^2\bc+\cosh ^2\bs+\sinh
   ^2 \bd\right)}  \,,
   \\
\nonumber
Y_{+} &=\frac{4 e^{-\bd } \cosh \bc \cosh \bs+2 \cosh 2
   \bc+e^{-2\bd}+1}{4 e^{-\bd } \cosh \bc \cosh
   \bs+e^{-2\bd} (2 \cosh 2 \bs+1)+1} \,,
    \\
\la{Yinfty}
Y_{-} &=e^{-2\bd}Y_{+} \,.
\end{align}
}

It is not difficult to see that the same formulae also describe the large $v$ asymptotics of the Y-functions of the B-model for any $\uh$ because at large $v$ Y-functions asymptote to constants and therefore the simplified TBA equations (\ref{stbay}-\ref{stbayw}) reduce to eqs.(\ref{u0TBAp}-\ref{u0TBAw}) with $\Delta$ replaced by $-2$. Thus the asymptotic values of Y-functions are  given by \eqref{Yinfty} with $\bd = -2\beta$. Furthermore these formulae determine the large $v$ asymptotics of the extended A- and B-models with $\Delta=-2\gh$
for the A- and opposite A-models, and $\Delta=-2-2\gh$
for the B-model and $\Delta=2-2\gh$ for the opposite B-model.

\subsection{{\it{t-J}} model from large on-site repulsion}\la{tJB}

In this appendix we consider the effect of large on-site Coulomb repulsion on the A-model and extended B-model. 
To this end it is useful to present the Hamiltonians of the models in the  form
  \begin{align}\notag
  {\mathbf H}(\g,\zeta,\kappa)&= \sum_{j=1}^L\, \Big( {\mathbf T}_{j,j+1} +4\g\, {\mathbf V}^H_{j,j+1}
  		+\zeta\,\big({\mathbf V}^H_{j,j+1}+  {\mathbf V}_{j,j+1}^{SS} -{\mathbf V}_{j,j+1}^{CC} + {\mathbf V}_{j,j+1}^{PH} \big) \Big)\,,\\ \la{hamAB}
  {\mathbf T}_{j,k}&= - \sum_{\sigma} \big(\cd_{j,\sigma} \cm_{k,\sigma}+\cd_{k,\sigma} \cm_{j,\sigma}\big)
  \Big(1- \big(1-\kappa_j\big) \big(\n_{j,-\sigma} - \n_{k,-\sigma}  \big)^2\Big)\,,
  \end{align}  
 where for the extended B-model  $\g=\gh$, $\zeta = 2\tanh \nu$
and $\ka_j=1/\cosh\nu$, while  for the A-model  $\g=\cosh\nu$, $\zeta = -2/\cosh\nu$
and $\ka_j=(-1)^j\tanh \nu$.
  
This form of the A-model Hamiltonian is obtained  by applying the unitary transformation \eqref{ut} to the Hamiltonian \eqref{HA2} with $\a=\pi/2$. 
Notice that as a result of the transformation  the charge $\su(2)$ generators take the same form as the ones for the Hubbard and B-models \eqref{su2s}.  
The correlated hopping term however becomes site dependent.

In the large $\g$ limit the Hubbard term dominates and the models are half-filled. It is useful to shift the chemical potential   
$\mu\to\mu+2\g$ to study the less than half-filled models, as this will remove the term proportional to ${\mathbf N}$ from the Hubbard interaction \eqref{VH}. 
Then let us write the Hamiltonian \eqref{hamAB} as ${\mathbf H}(\g,\zeta,\kappa) =4 \g\sum_{j=1}^L \n_{j,\uparrow} \n_{j,\downarrow}  +   {\mathbf H}(0,\zeta,\kappa)$ and note that
in the strict $\g\rightarrow \infty$ limit any state that does not contain doubly occupied sites is a ground state and there is huge degeneracy. 
Taking the sub-leading terms into account in $\frac{1}{\g}$ perturbation theory in a standard way, see e.g. section 2.A and particularly equations (2.A.26, 2.A.30) of the book \cite{book}, the effective large $\g$ Hamiltonian is
\bean
  {\mathbf H}^{t-J}(\g,\zeta,\kappa) ={\mathbf P}_0\,  {\mathbf H}(0,\zeta,\kappa)\, {\mathbf P}_0 +
			\frac{1}{4 \g}\sum_{j=1}^{L} {\mathbf P}_0 \,{\mathbf H}(0,\zeta,\kappa)\,   \n_{j,\uparrow} \n_{j,\downarrow}  \,{\mathbf H}(0,\zeta,\kappa) \,{\mathbf P}_0  \,,
\eean
where ${\mathbf P}_0=\prod_{j=1}^L(1- \n_{j,\uparrow} \n_{j,\downarrow} )$ is the projector onto the model's Hilbert space of dimension $3^L$ where there are no doubly occupied sites. As happens for the similar analysis of the Hubbard model, the second term in the above expansion gives rise to three-site terms as well as two-site terms. The three site-terms however will be suppressed in the physically interesting regime where the model is close to half-filling and so we will ignore them here, as is generally done.

 The effective Hamiltonian is thus
\bean
  {\mathbf H}^{t-J}(\g,\zeta,\kappa)= {\mathbf P}_0 \sum_{j=1}^{L} \Big(-  \cd_{j,\sigma} \cm_{j+1,\sigma}- \cd_{j+1,\sigma} \cm_{j,\sigma} 
  	+( \zeta + {\kappa^2 \ov  \g } )\big({\mathbf V}_{j,j+1}^{SS}  - \frac{ \n_{j} \n_{j+1}}{4}\big) \Big)  {\mathbf P}_0\,,
\eean
which describes the  {\it t-J}  model with the coupling $J= \zeta + {\kappa^2 \ov  \g } $.
For the A-model expanding $J$ in powers of $1/\uh$ one gets $J=-1/\uh$, and therefore  a ferromagnetic \tj model.
For the B-model  $J=2  \tanh \nu + {1 \ov  \gh \cosh^2 \nu}$, and at $\nu=\infty$ it is the supersymmetric \tj model.


\end{document}